\documentclass[a4paper,man,floatsintext,natbib]{apa6}

\def\pprw{8.5in}
\def\pprh{11in}

\usepackage{setspace}

\usepackage[english]{babel}
\usepackage{booktabs,times}
\usepackage[utf8x]{inputenc}
\usepackage{amsmath}
\usepackage{graphicx}
\usepackage[colorinlistoftodos]{todonotes}
\usepackage{authblk}
\usepackage{balance}

\begin{document}

\onehalfspacing

\title{Quantifying Biases in Online Information Exposure}
\shorttitle{Quantifying Biases in Online Information Exposure}
\author[*]{Dimitar Nikolov}
\author[+]{Mounia Lalmas}
\author[*]{Alessandro Flammini}
\author[*]{Filippo Menczer}
\affil[*]{Center for Complex Networks and Systems Research, Indiana University}
\affil[+]{Yahoo Research @ Oath}
\affiliation{~}

\abstract{Our consumption of online information is mediated by filtering, ranking, and recommendation algorithms that introduce unintentional biases as they attempt to deliver relevant and engaging content. It has been suggested that our reliance on online technologies such as search engines and social media may limit exposure to diverse points of view and make us vulnerable to manipulation by disinformation. In this paper, we mine a massive dataset of Web traffic to quantify  two kinds of bias: (i)~homogeneity bias, which is the tendency to consume content from a narrow set of information sources, and (ii)~popularity bias, which is the selective exposure to content from top sites. Our analysis reveals different bias levels across several widely used Web platforms. Search exposes users to a diverse set of sources, while social media traffic tends to exhibit high popularity and homogeneity bias. When we focus our analysis on traffic to news sites, we find higher levels of popularity bias, with smaller differences across applications. Overall, our results quantify the extent to which our choices of online systems confine us inside ``social bubbles.''}

\maketitle

\section{Introduction}
\label{sec:intro}

Our online information ecosystem has seen an explosive growth in the number of information producers, consumers, and content.
Web platforms have become the dominant channels that mediate access to information in this ecosystem.
For example, users rely on social media for their news consumption more than ever before~\citep{Pew2017}.
To help us cope with information overload, Web platforms use algorithms to steer our attention toward important, relevant, timely, and engaging content.
For instance, search engines use variations of PageRank to estimate the global reputation of Web pages and rank results accordingly~\citep{Google}; recommendation systems leverage the similarities between items and among user preferences~\citep{AmazonRec,GoogleNewsRec}; and online social networks nudge users to consume content shared by their friends~\citep{FacebookFeed}.

The biases embedded in these information filtering algorithms may have unintended consequences. 
Reliance on popularity signals such as PageRank, trending topics, likes, and shares may lead to an entrenchment of established sources at the expense of novel ones~\citep{Hindman,Cho}.
Dependence on engagement metrics may also make us vulnerable to manipulation by orchestrated campaigns and social bots~\citep{astroturf,social-bots}.
Exposure to news through the filter of the social network of like-minded individuals may bias our attention toward information that we are already likely to know or agree with.
So-called ``echo chambers''~\citep{Sunstein3} and ``filter bubbles''~\citep{Pariser} have been claimed to be pathological consequences of social media algorithms and to lead to polarization~\citep{Truthy,Sunstein0}.
Homogeneous social groups facilitated by online interactions may also make people vulnerable to misinformation~\citep{mckenzie2004,stanovich2013,Jun06062017}.

Given the impact of Web platforms on our information consumption, it is critical to understand and quantify two kinds of algorithmic bias alluded to above: \emph{homogeneity bias}, which we define as the selective exposure of content from a narrow set of information sources; and \emph{popularity bias}, which is the tendency to expose users to content from popular sources.
Note that since we focus on the algorithmic biases of platforms, our definitions are based on \emph{exposure} --- a platform exposes users to information sources in a biased way.
However, we measure exposure by observing the \emph{sources} of content that are actually \emph{consumed} by the users. This operational definition is based on the fact that we do not have data about what pages users see, but only about what pages they visit. 
Armed with these definitions, we investigate three hypotheses in this paper:

\begin{itemize}
\item {\bf H1}: All Web platforms have some popularity and homogeneity bias, but might differ greatly in how much.
\item {\bf H2}: Social media platforms exhibit higher homogeneity bias compared to other platforms due to homophily in social network.
\item {\bf H3}: Search engines exhibit higher popularity bias compared to other platforms due to their reliance on PageRank-like metrics.
\end{itemize}

H1 is at the heart of the concerns raised by Pariser and Sunstein, with a key question being the extent to which the biases displayed by different platforms deviate from a baseline. Regarding H2, in prior work~\citep{Nikolov}, we have already found evidence in support of this hypothesis, but only by aggregating information consumption accross users, and studying broad categories of activity rather than specific platforms. The present study extends those prior results by analyzing individual users, using more recent data, and comparing individual platforms. Evidence for H2 is also found in existing research on social media platforms and blogs (see ``Background''), but the present study is the first to compare bias across different platforms. Finally, regarding H3, search engines have been found to mitigate the popularity bias that stems from the scale-free structure of the Web graph because of their ability to respond to specific and heterogeneous queries~\citep{Fortunato}. However, this finding was published more than ten years ago. How popularity bias in online systems has evolved with these changes is an open question. 

In summary, this paper makes the following contributions:
\begin{itemize}
  \item We \emph{formally define}, \emph{measure}, and \emph{compare} homogeneity and popularity biases in several widely used Web platforms in five categories: email, online social networks, search engines, news recommenders, and Wikipedia. These categories represent a significant portion of online traffic and allow us to indirectly observe the effects of several distinct algorithms and mechanisms: retrieval, ranking, crowdsourcing, communication, and personalization. We apply these measures to click data, which capture information that is consumed by users; the same measures can be applied to different activities, such as social sharing.
  \item In support of H1, we show that all investigated platforms are biased compared to a random baseline, but we find quantitative differences in bias measurements.
  \item In support of H2, aggregate traffic analysis reveals that social media exhibit higher homogeneity bias than other platforms. However, they also exhibit higher popularity bias.
  \item We refine previous results about the relationship between search traffic volume and domain rank; we find that user attention is biased toward popular websites according to the rich-get-richer structure of the Web. However, contrary to H3, search engines exhibit less popularity bias compared to other platforms. When we focus our analysis on traffic to news sites, we find higher levels of popularity bias.
  \item We further investigate the biases associated with individual traffic patterns and report on the correlation between homogeneity and popularity bias across systems.
\end{itemize}

The rest of this paper is structured as follows. In Section ``Background,'' we examine related work on measuring Web biases and their consequences. In Section ``Dataset,'' we give the details of the Web traffic data analyzed here. Section ``Bias Measures'' defines the homogeneity and popularity bias measures, and in Section ``Results'' we apply these metrics to a variety of platforms. In the ``Discussion'' section, we put our findings in more general context and discuss further work. In the ``Appendix,'' we present results from different normalizations of the measures to demonstrate their robustness.

\section{Background}
\label{sec:related}

The study of exposure to diverse sources of information in online systems is often motivated by societal questions, such as polarization of opinions related to news and politics. Many studies have focused on specific systems. For example, among bloggers, the existence of echo chambers can be fostered by liberal and conservative users who link primarily within their own communities~\citep{Adamic}, and commenters who are more likely to agree with each other than to disagree~\citep{Gilbert}. On Facebook, users are more likely to share news articles with which they agree~\citep{An1,An2,An3,Grevet}. Also on Facebook, three filters --- the social network, the news feed algorithm, and a user's own content selection --- significantly decrease exposure to ideologically challenging news~\citep{BakshyAdamic}. More generally, we are somewhat more likely to be exposed to news articles aligned with our worldviews when we browse online~\citep{Flaxman}. These effects, however, are not static --- the way in which news are presented can affect the willingness of users to engage with a piece of information with which they disagree~\citep{DorisDown,Munson,GraellsGarrido1,GraellsGarrido2}. Exposing the choices of others may also lead to homogeneity bias~\citep{Salganik}. All these studies lend credence to H2 from the Introduction --- that social media tend to expose users to content from a homogeneous set of sources.

At the same time, there is some empirical evidence suggesting that, contrary to H2, some platforms have the potential to diversify the information to which we are exposed.
For example, recommendations based on users who select similar sets of items may lead to heterogeneous choices~\citep{Hosanagar,Fleder}. The present work does consider one recommendation system, Google News, but does not directly examine the recommendation effect on exposure. On Facebook, \citep{BakshyAdamic} found that despite decreased exposure, users still see about 25\% of ideologically challenging news on their feed. Another study showed that the communities around news media's Twitter accounts post less prominent stories than the media they follow~\citep{SaezTrumper2013}. This finding can be interpreted as showing that social media can serve to diversify the stories we read with more niche ones. However, these niche stories can contribute to an echo chamber effect if they are the only ones consumed by a user, since by definition they are of interest to a narrower audience; the same study found that although users may take interest in a broad range of stories, they focus most of their attention on a few.

The popularity bias of Web search engines has received considerable attention in the literature. Consistent with H3, it has been argued that search engines amplify the rich-get-richer effect of the Web graph: popular Web pages would attract even more visitors than they would without search engines, yielding a superlinear relationship between traffic and PageRank~\citep{Cho,Introna,Mowshowitz}. Empirical results have not supported this hypothesis, showing instead that popular websites receive less traffic than predicted by a rich-get-richer model~\citep{Fortunato}. The heterogeneity of user queries provided a quantitative interpretation for this finding. However, there have not been other empirical studies on this question, and search patterns and algorithms may have changed in the ten years since, making it necessary to revisit the question of popularity bias empirically.

In previous work we analyzed Web traffic aggregated across users, finding that sources reached through social media as a whole are significantly less diverse than those exposed via email and search engines~\citep{Nikolov}.
In this paper, we follow a similar methodology of considering Web click data without inspecting the content of the visited sites. While the studies cited above focus on specific events or US political news consumption, our content-independent approach is applicable to a wide variety of topics, geographical regions, and languages. In addition, traffic data allow us to consider a wider range of users, and to capture a wider range of engagements for each user, compared to previous studies.
We extend our previous methodology by (i) analyzing individual users, rather than aggregate data, (ii) considering a more up-to-date dataset, (iii) measuring popularity bias in addition to homogeneity, and (iv) comparing individual platforms.

\section{Dataset}
\label{sec:dataset}

To study bias in information exposure, we used a large Web traffic dataset of Yahoo Toolbar users collected between July 1, 2014 and March 31, 2015. The dataset contains anonymized browsing data, consisting of \texttt{(timestamp, browser cookie, target domain, referring domain)} tuples. The users in the sample gave their consent to provide data through the Yahoo Toolbar.

We considered a large sample of the most popular referrers in the dataset from the five categories of interest seen in Table~\ref{tbl:referrers}. The sample consists of nearly \textbf{2.4 billion clicks}
%2,388,846,925
from over \textbf{10.5 million unique users}
% 10,582,851
to over \textbf{33.7 million unique targets}.
% 33,745,639

\begin{table}
\centering
\caption{Categories and referrers of interest.}
\begin{tabular}{ll}
  \hline
  \textbf{Category} & \textbf{Referrers}\\
  \hline
  Email & Yahoo Mail, GMail, AOL Mail\\
  Social Media & Facebook, Twitter, Reddit, YouTube, Tumblr, Pinterest\\
  Web Search & Yahoo Search, Google, Bing, Ask\\
  News recommendation & Google News\\
  Wiki & Wikipedia\\
  \hline
\end{tabular}
\label{tbl:referrers}
\end{table}

Studying traces of Web traffic has advantages over content-based approaches. First, it allows us to compare several different Web platforms. Second, a click has a lower cognitive cost than behaviors such as writing blog entries or Twitter posts. This enables the collection of large volumes of user actions that represent a wide set of online experiences and decisions. On the other hand, content inspection would permit a more detailed analysis of the information to which users are exposed. Thus, an important next step in the study of exposure bias is to verify the present findings when taking content into account. Furthermore, Yahoo users may be a biased sample of the population of Internet users, and this may correspond to a bias in our sample of traffic. Other studies have used similar dataset when analyzing Web browsing patterns~\citep{DBLP:journals/jasis/LehmannCLB17,west2012drawing}. Nevertheless, as with any observational study of Web users, results should be reproduced in different populations.

In the dataset described thus far, we have constrained the sources of clicks to the platforms in Table~\ref{tbl:referrers}, but the targets are unconstrained. However, in the context of echo chambers and polarization, we are particularly interested in examining traffic toward news sites because divisive political discourse often occurs around news stories. Thus we created a separate dataset consisting only of clicks to news website domains. The list of possible targets was created by traversing relevant categories in the Open Directory (\url{dmoz.org}) using a procedure described in the literature~\citep{Nikolov}. The list of news targets includes a broad range of news providers, including for example local newspapers, rather than focusing on popular sources only. This procedure resulted in % 3,238
over \textbf{3,200 news sites}. We used this list to filter the targets in the click collection, yielding the news traffic dataset referenced in the ``Results'' Section. 

\section{Bias Measures}
\label{sec:methods}

In this section we formally define homogeneity and popularity bias and discuss different ways of sampling clicks from the dataset.

\subsection{Homogeneity Bias}

\emph{Homogeneity bias} is the tendency of an application (e.g., platform or website) to generate traffic disproportionally to a small set of target websites. For example, consider 10 clicks originating from the same Web application and landing on the domains $D_1$, $D_2$ and $D_3$: the highly-skewed distribution of clicks $\{D_1: 8, D_2: 1, D_3: 1\}$ has higher homogeneity bias compared to the more even distribution $\{D_1:4, D_2: 3, D_3:3\}$. We capture this intuition by using the \emph{normalized inverse Shannon entropy} to define the homogeneity bias experienced by user $u$ through application $a$: 
\[
B_h(u,a) = 1 + \frac{\sum_{t \in D(C_{u,a})} p_{t}\log(p_{t})}{\log(|D|)}
\]
where $C_{u,a}$ is a sample of clicks by $u$ through $a$, $D(C_{u,a})$ is the set of target domains reached through the clicks, $D$ is the set of all target domains, and $p_{t}$ is the fraction of clicks requesting pages from target domain $t$. The entropy is maximized when all outcomes are equally likely, yielding the normalization factor $\log(|D|)$. $B_h$ is thus defined in the unit interval; it is zero when traffic is distributed equally across all targets and one when it is concentrated toward any single domain. We measure the homogeneity bias of an application by averaging across users: \[B_h(a) = \langle B_h(u,a) \rangle_u.\]

\subsection{Popularity Bias}

\begin{figure}[t]
\centering
\includegraphics[width=0.7\textwidth]{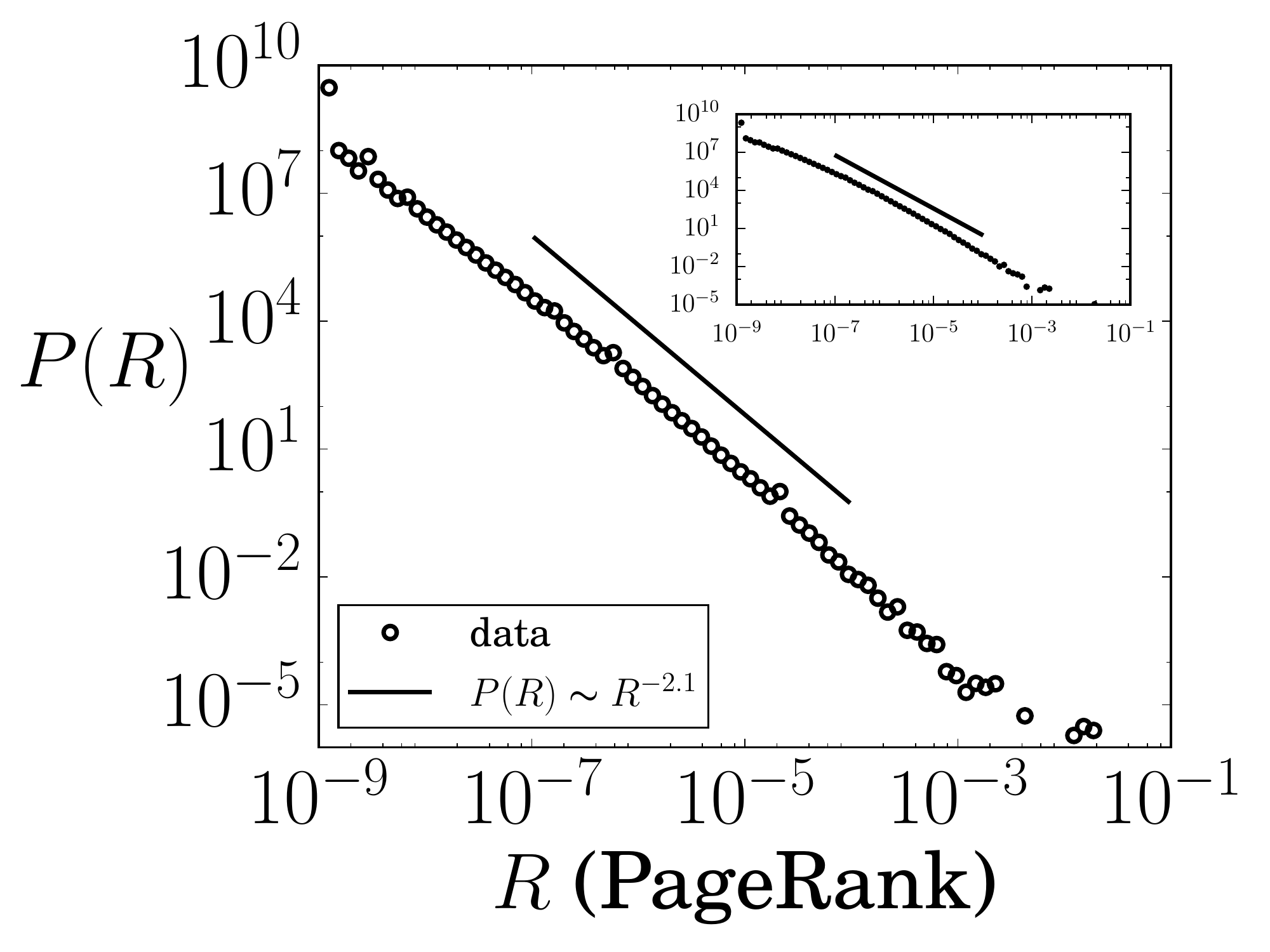}
\caption{PageRank distribution for the domain graph recovered from a large Web crawl. The inset shows the distribution of the domains in the Yahoo Toolbar data. The solid lines show the power law $P(R) \sim R^{-2.1}$.}
\label{fig:prdistro}
\end{figure}

\emph{Popularity bias} is the tendency of an application to generate traffic to already popular websites. \cite{Fortunato} measured popularity bias by comparing the traffic to a website against a null model based on PageRank~\citep{Google}. Since PageRank models a random walk over the network, it captures traffic patterns that reflect the scale-free structure of the underlying Web graph. Indeed, PageRank has the same power-law distribution as the in-degree, as shown in Figure~\ref{fig:prdistro} for our dataset, and in the literature~\citep{Fortunato08internetmath,Broder}. Therefore, the null model predicts that traffic should be proportional to PageRank ($T \sim R$).
%\todo{I changed this, is this still correct? Dimitar: Yes, PR and in-degree both follow a power-law with exponent ~2.1; the Fortunato08 paper shows this for PR for a number of crawls/Web graphs; for in-degree, this is shown by the Broder paper I added a citations to here.}

Measuring the scaling relationship between traffic and PageRank requires the fitting of traffic curves, which is sensitive to the range and amount of fitted data. For this reason, \cite{Fortunato} aggregated traffic across users and applications. We need a way to quantify the popularity bias experienced by an individual user when using a particular platform. Here, we propose a different measure that achieves these goals.

We follow the same tradition of measuring popularity by PageRank centrality, which we compute over the domain graph of over 200 million
% 201,833,342
domains with over 1.4 billion
% 1,428,652,887
links by the recursive function
\[
R(i) = \frac{\alpha}{N} + (1 - \alpha)\sum_{j \in L_i} \frac{R(j)}{N_j}.
\]
Here, $i$ and $j$ are domains, $N$ is the total number of domains, $\alpha$ is the so-called teleportation factor ($\alpha=0.15$), $L_i$ is the set of domains with links to $i$, and $N_j$ is the number of domains linked from pages in domain $j$. This definition of popularity is independent of the traffic data and therefore not subject to its bias. 

\begin{figure}[t]
\centering
	\includegraphics[width=0.32\textwidth]{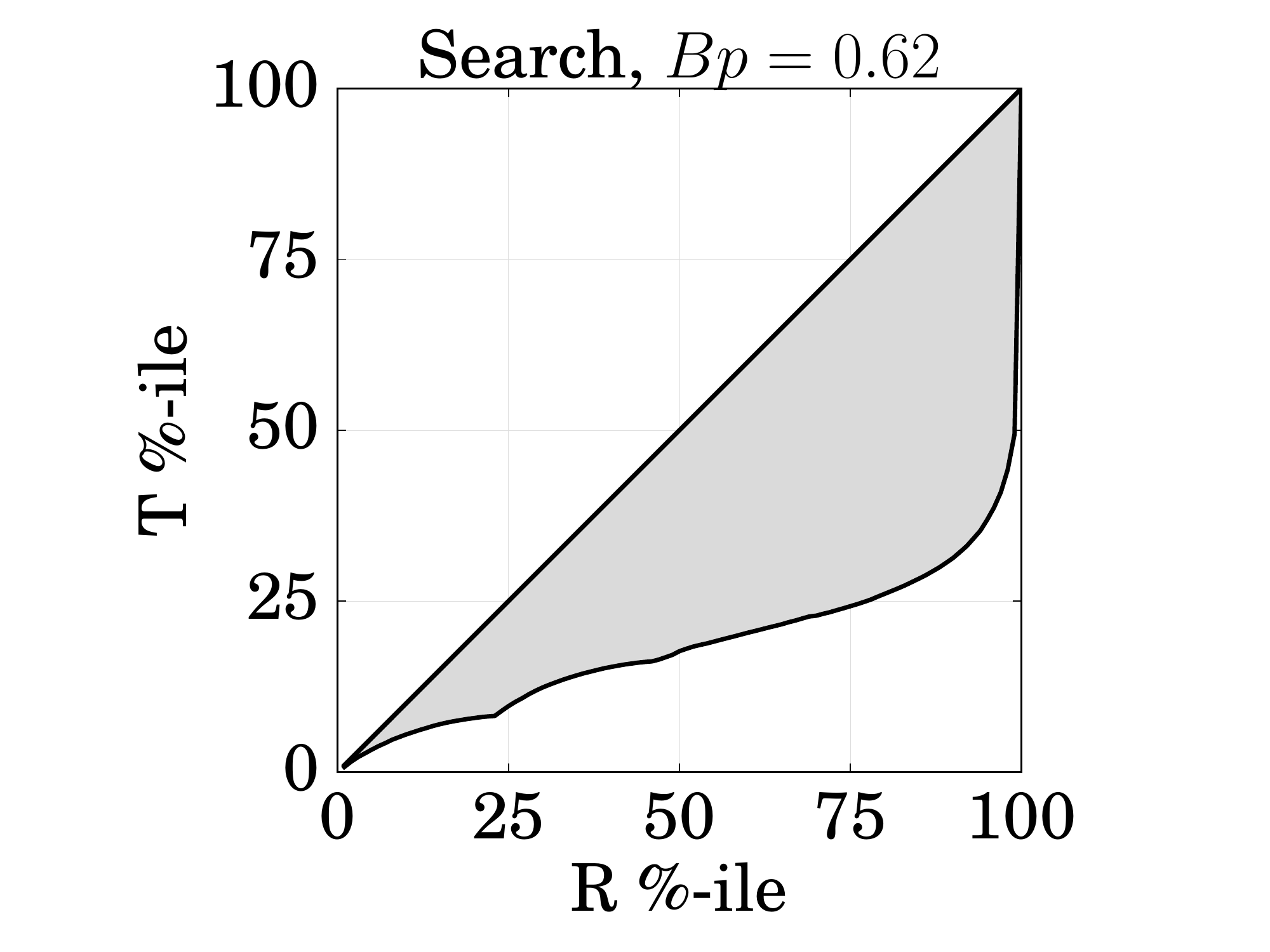}
	\includegraphics[width=0.32\textwidth]{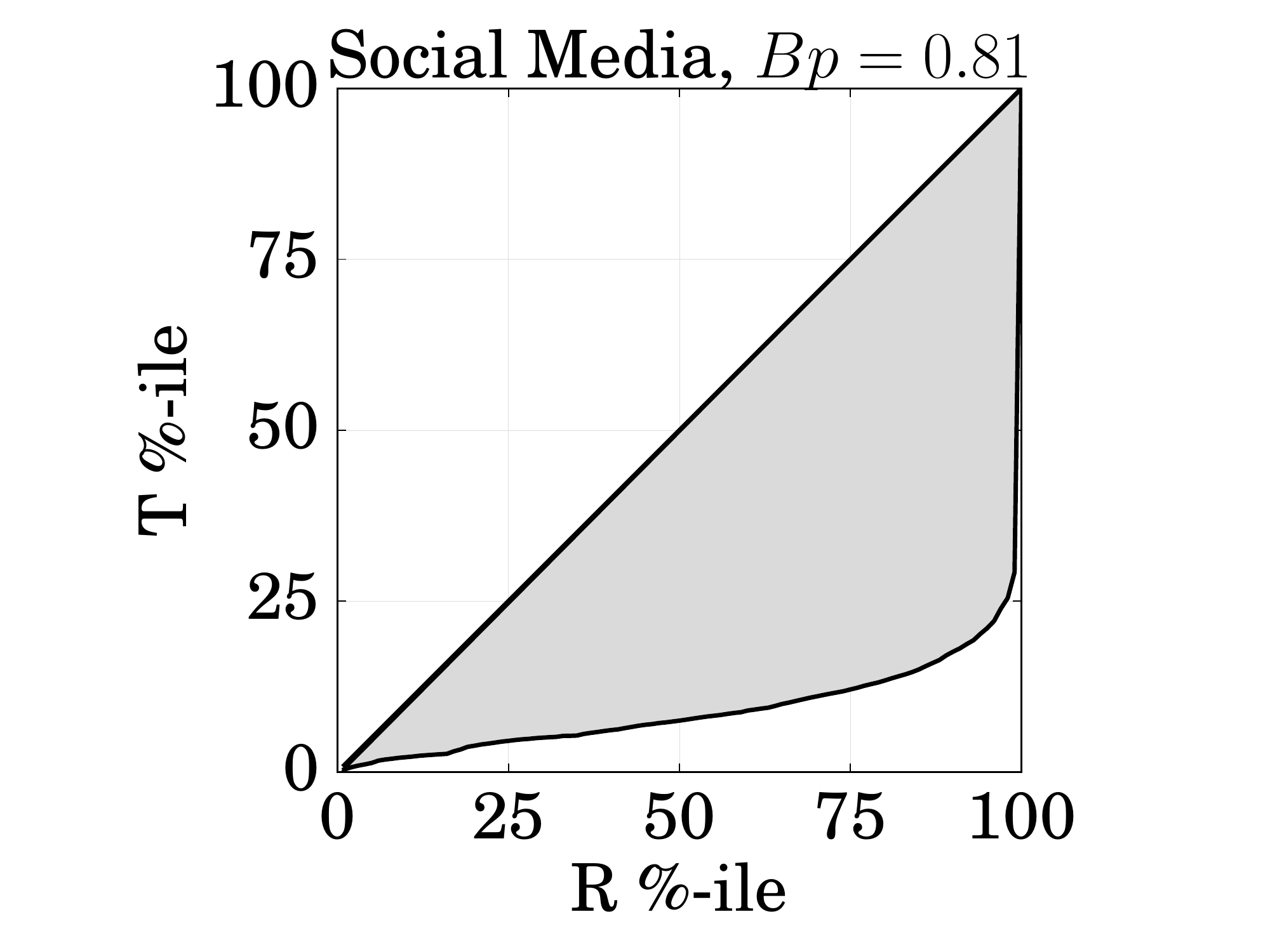}
	\includegraphics[width=0.32\textwidth]{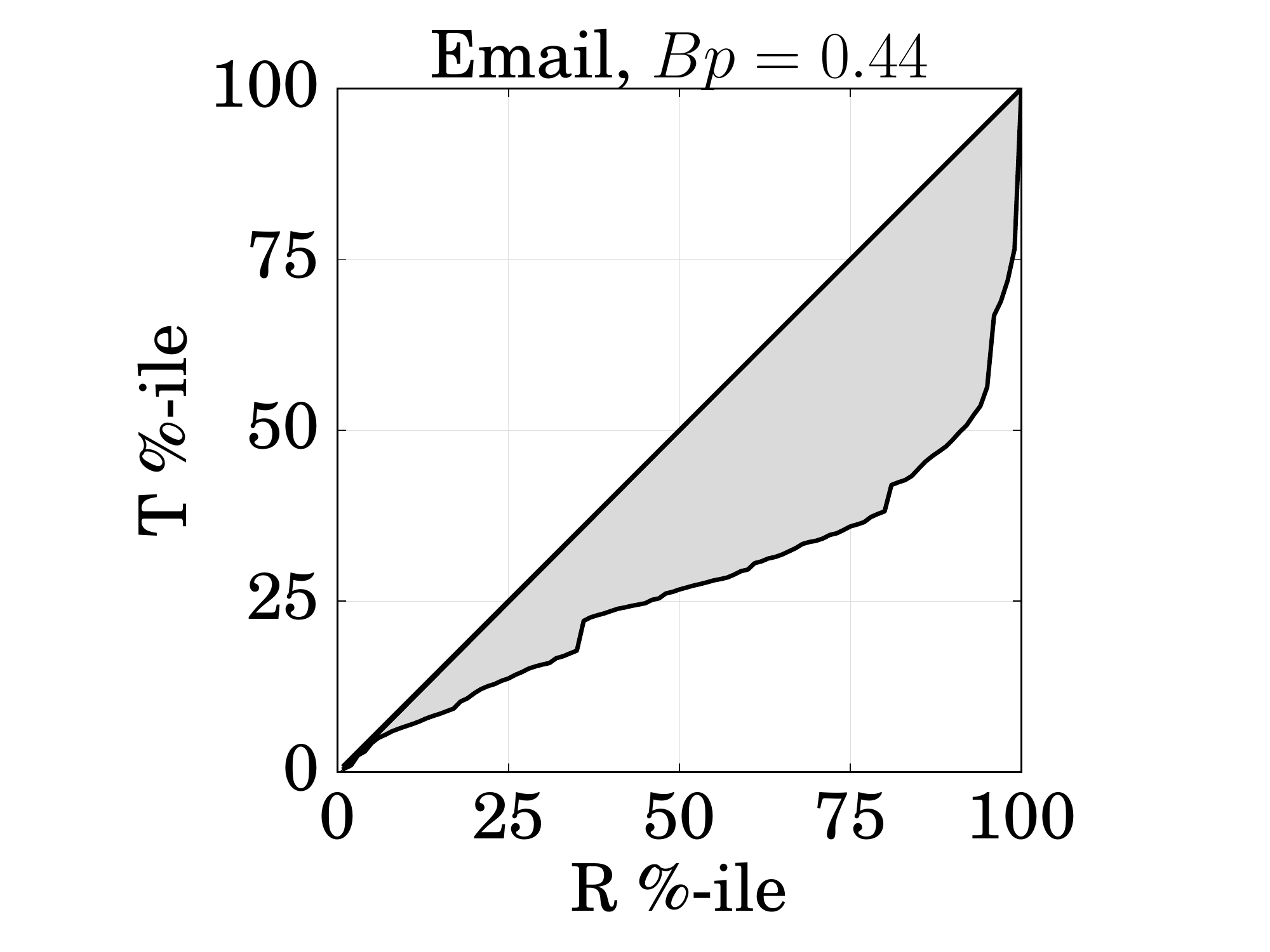}\\
	\includegraphics[width=0.32\textwidth]{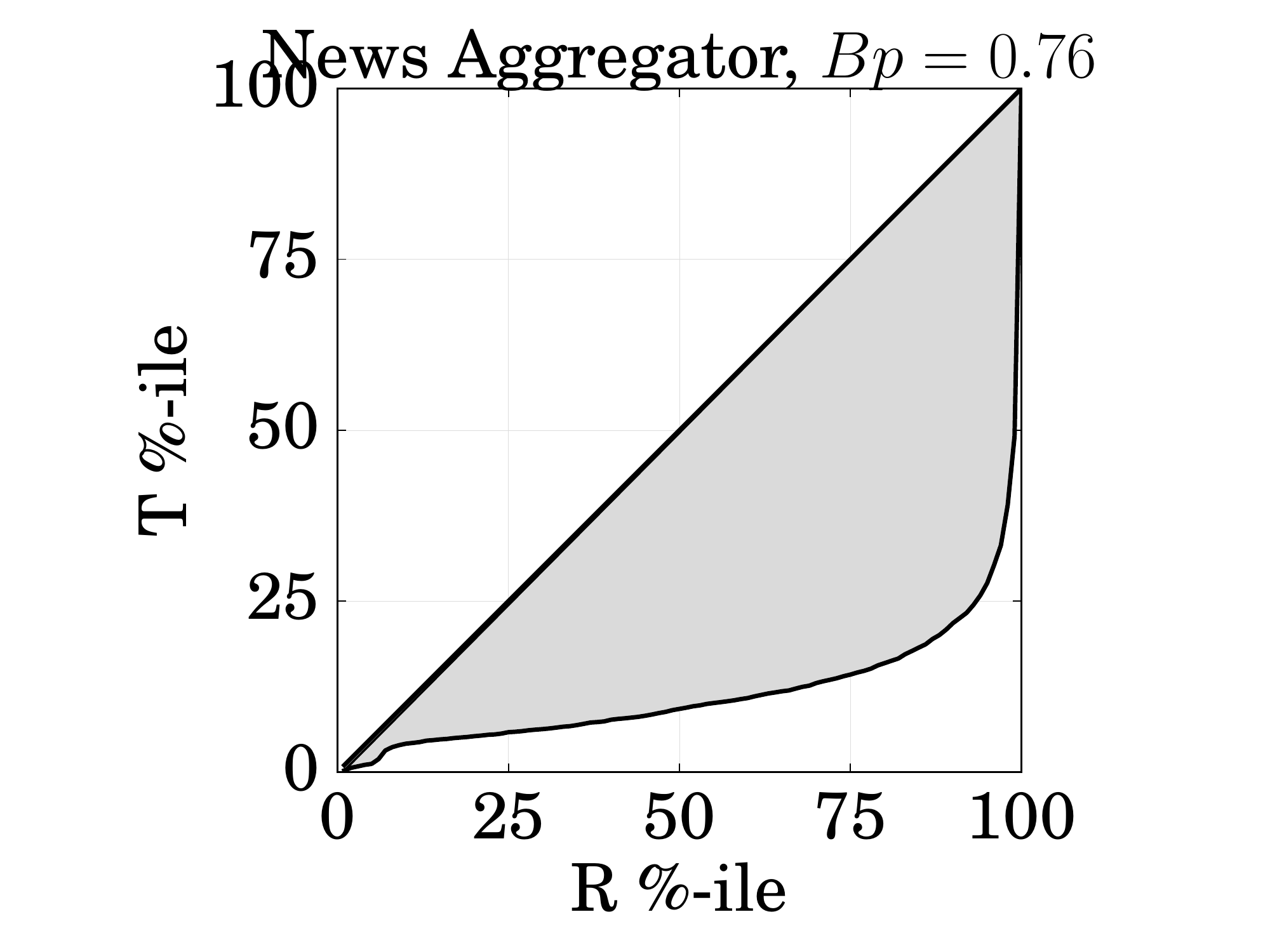}
	\includegraphics[width=0.32\textwidth]{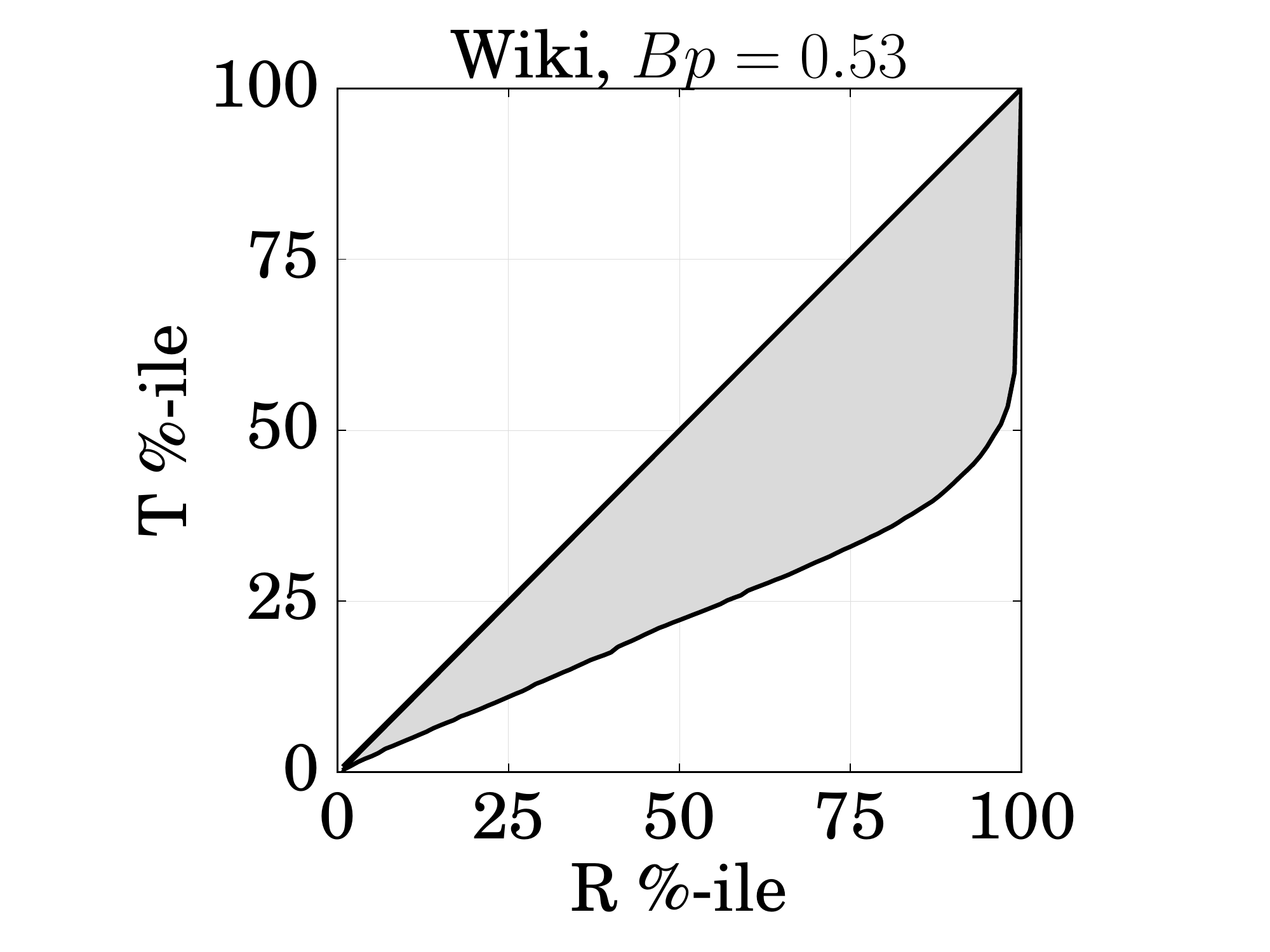}
\caption{Lorenz curves of traffic $T$ versus PageRank $R$ show the popularity bias of a random user for each category. The curves are used to measure the Gini coefficient (popularity bias), illustrated by the shaded areas. The diagonals show equal distributions, where traffic is proportional to PageRank. The actual distributions show that targets with low PageRank receive less traffic, while targets with high PageRank receive the vast majority of the traffic, suggesting bias towards globally popular domains.}
\label{fig:qqcats}
\end{figure}

To calculate the popularity bias of an application $a$ experienced by user $u$, we use the Lorenz curve~\citep{lorenz} of the cumulative traffic volume $V_{u,p}(r)$ as a function of PageRank percentile $r$. Figure~\ref{fig:qqcats} show several Lorenz curves computed for the same random user as an illustration of this concept. A line below the diagonal signifies a bias in favor of more popular sites.

Let us therefore define popularity bias as the area between the Lorenz curve and the diagonal. This quantity is expressed by the Gini coefficient:
\[
B_p(u, a) = 1 - 2\int_{0}^{1} V_{u, a}(r) dr
\]
where the traffic share $V$ is measured based on a sample of clicks. $B_p$ is zero when the traffic is independent of the target popularity and one when it is concentrated toward the domain with the highest PageRank.
%; $B_p$ can be negative if traffic is concentrated on  domains with low PageRank. 
The popularity bias of an application is obtained by averaging across users: \[B_p(a) = \langle B_p(u,a) \rangle_u.\]

\subsection{Baseline Biases}

To gauge our measurements of popularity and homogeneity bias, we consider a null model of traffic as a baseline. In this model, we simulate a random walker on the Web traffic graph, modified with a 15\% teleportation probability, and with the probability of following links proportional to their traffic. This random walk process is a modified version of PageRank that takes into account the weights of the links~\citep{Meiss2010WAW}. We simulate the same number of random walkers as sampled users, and the same number of steps as the number of clicks sampled for each user. We then measure the homogeneity and popularity biases from the resulting traces through the websites. We report these baselines in all experiments. Our use of the modified PageRank as a baseline is an attempt to
account for the benefit of some amount of bias toward popular sites, which may have acquired popularity due to higher quality, trust, prestige, or other desirable properties.

\subsection{Traffic Sampling}

Because our goal is to compare different applications, we want the sample of clicks $C_{u,a}$ to represent equal efforts by users across applications. There is no single best way to accomplish this goal when sampling a user's clicks, so we used two different methods. One method is to uniformly sample the same number of clicks for each $(u,a)$ pair, thus equating effort with click volume. This approach is motivated by volume effects in measures of traffic heterogeneity~\citep{Nikolov}. The size of our dataset allowed us to sample 500 users per individual system, and 1,200 users per aggregated category with at least 100 clicks each. For the news dataset, there were fewer users with the desired number of clicks. We excluded applications with fewer than 30 users from our analysis. An alternative approach is to collect all clicks in a fixed time period, thus equating effort with time. In the ``Results'' section we report the findings from sampling the same number of clicks. In the Appendix we have included results from sampling during a fixed time period; the results are consistent.

One drawback to both sampling approaches is that there are few users with sufficient traffic from \emph{all} systems under consideration. Therefore, it is not possible to have a single sample of users to measure the biases for all applications. Instead, we sample a separate group of users with sufficient traffic for each application. To test the robustness of this method, we considered a sample of users with enough traffic in each of the three top categories (\emph{search}, \emph{social media} and \emph{email}) and found the aggregate bias measurements to be consistent, as we show in the ``User Analysis'' Subsection.

\section{Results}
\label{sec:results}

Using the measures we have defined in the previous section, we can now quantify the exposure biases in our large dataset of clicks.

\subsection{Homogeneity Bias}

\begin{figure}[t]
\centering
\begin{tabular}{cc}
A & B\\
\includegraphics[width=0.49\textwidth]{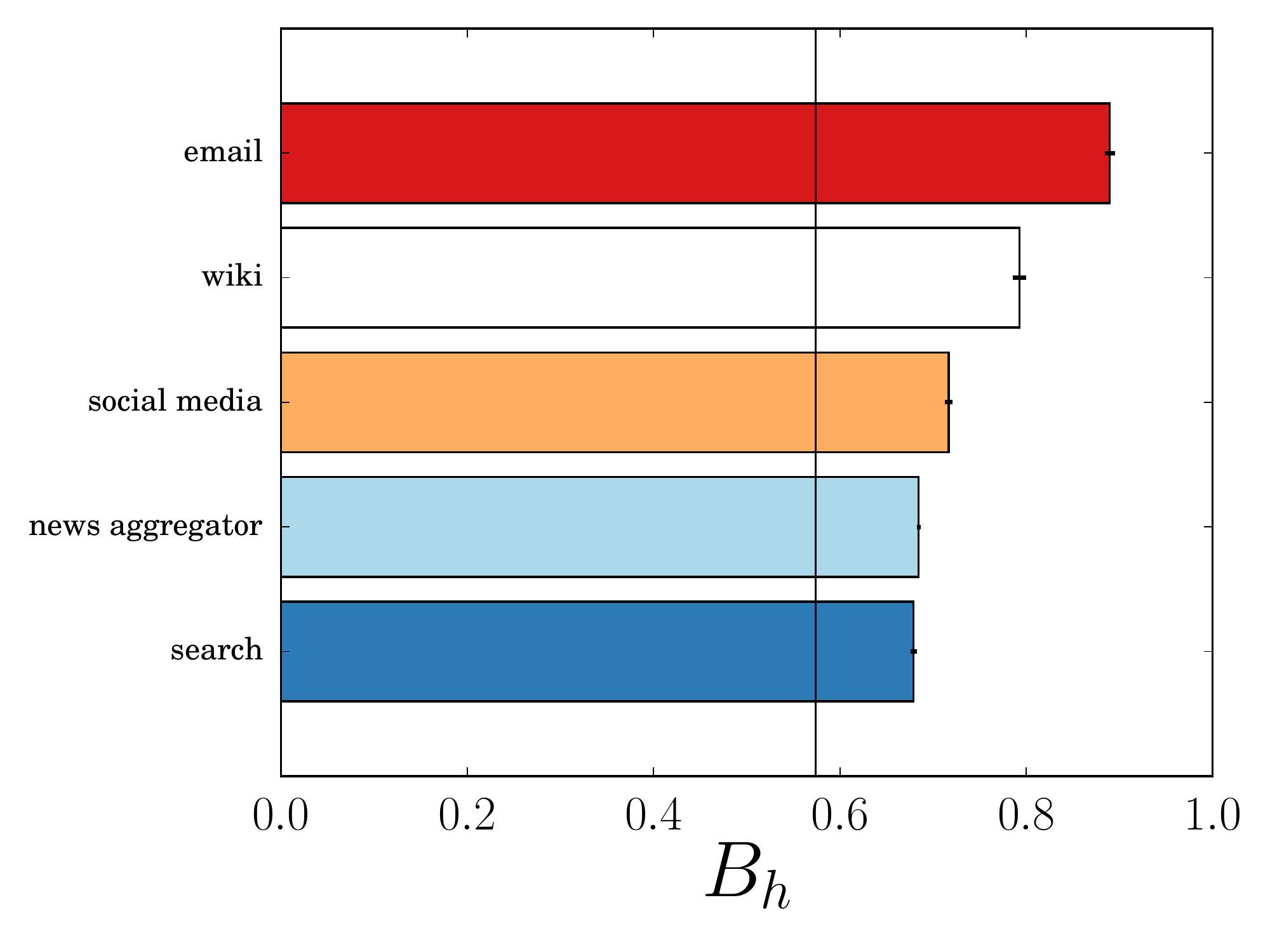} &
\includegraphics[width=0.49\textwidth]{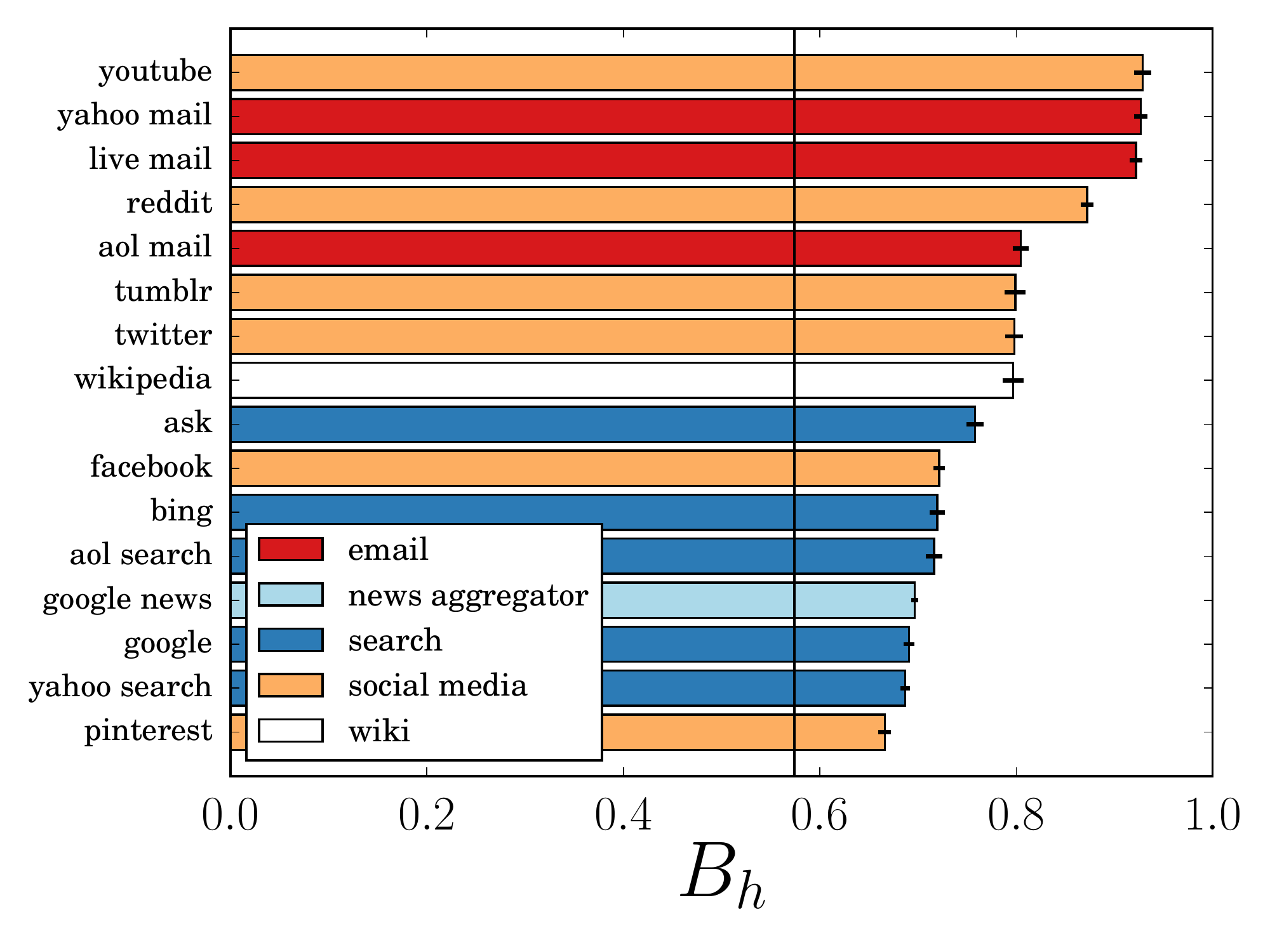} \\
\end{tabular}
\caption{Homogeneity bias for (A) aggregate categories of interest and (B) individual domains. Error bars are $\pm 2$ standard errors. The vertical lines show the baseline bias of a random walker through the Web domain network.}
\label{fig:hbias}
\end{figure}

Recall that homogeneity bias is defined as the tendency of a platform to expose users to information from a narrow set of sources. The homogeneity bias results are shown in Figure~\ref{fig:hbias}. We observe the highest homogeneity bias for email and lowest for search, as seen in Figure~\ref{fig:hbias}A. In addition, all platforms are more biased than the baseline (Figure~\ref{fig:hbias}B), and there are significant differences between social media platforms, suggesting that they are better examined separately rather than in aggregate. These findings support H1. Facebook, which dominates social media traffic has moderate homogeneity bias, putting the combined category in the middle and resulting in mixed evidence for H2. Pinterest displays the lowest homogeneity bias, suggesting that this platform exposes users to a broad range of different information sources.

\subsection{Popularity Bias}

\begin{figure}[t]
\centerline{\includegraphics[width=0.7\textwidth]{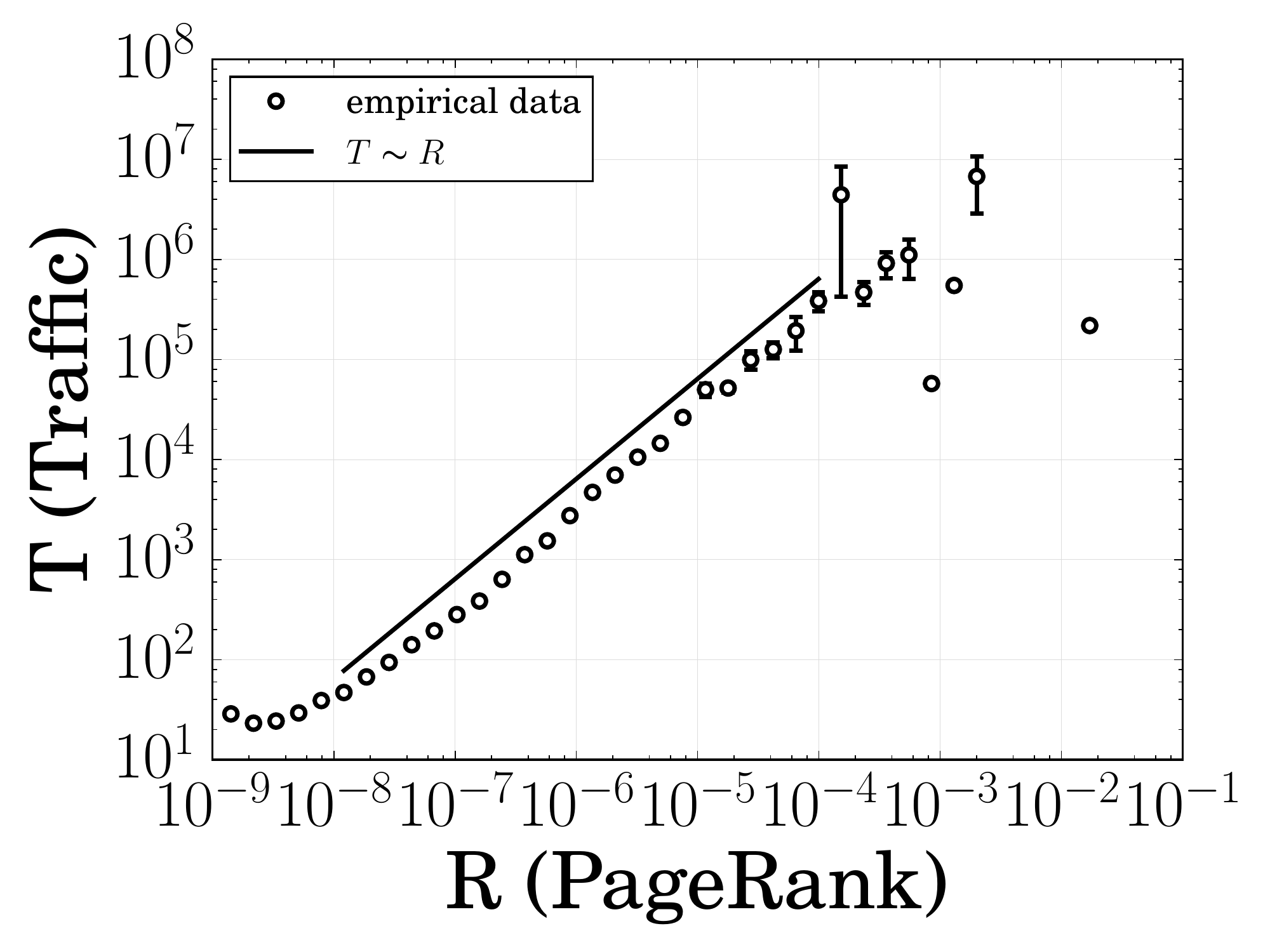}}
\caption{Traffic volume to a domain versus its PageRank. The line shows a proportionality relationship  between the two.} 
\label{fig:pr-vs-volume-all}
\end{figure}

Recall that popularity bias is defined as the tendency of a platform to expose users to information from popular sources. \cite{Fortunato}'s dataset was dominated by search traffic, and their analysis found that popular websites receive less traffic $T$ than predicted by the PageRank null model ($T \sim R^{\gamma}$ with $\gamma < 1$), attributing this sublinear behavior to search queries reflecting specific user interests. Measuring this relationship more than ten years later allows us to investigate how the popularity bias of the Web has evolved since then. 
The scaling relationship between traffic volume and PageRank in our dataset is shown in Figure~\ref{fig:pr-vs-volume-all}. We find a strong linear relationship $(\gamma = 1)$ over four orders of magnitude. This indicates an overall increase in bias compared to 2006. However, we find no support for the previously claimed superlinear behavior ($\gamma > 1$)~\citep{Introna,Mowshowitz,Cho} that would be consistent with H3.

\begin{figure}[t]
\centering
\begin{tabular}{cc}
	A & B \\
 	\includegraphics[width=0.49\textwidth]{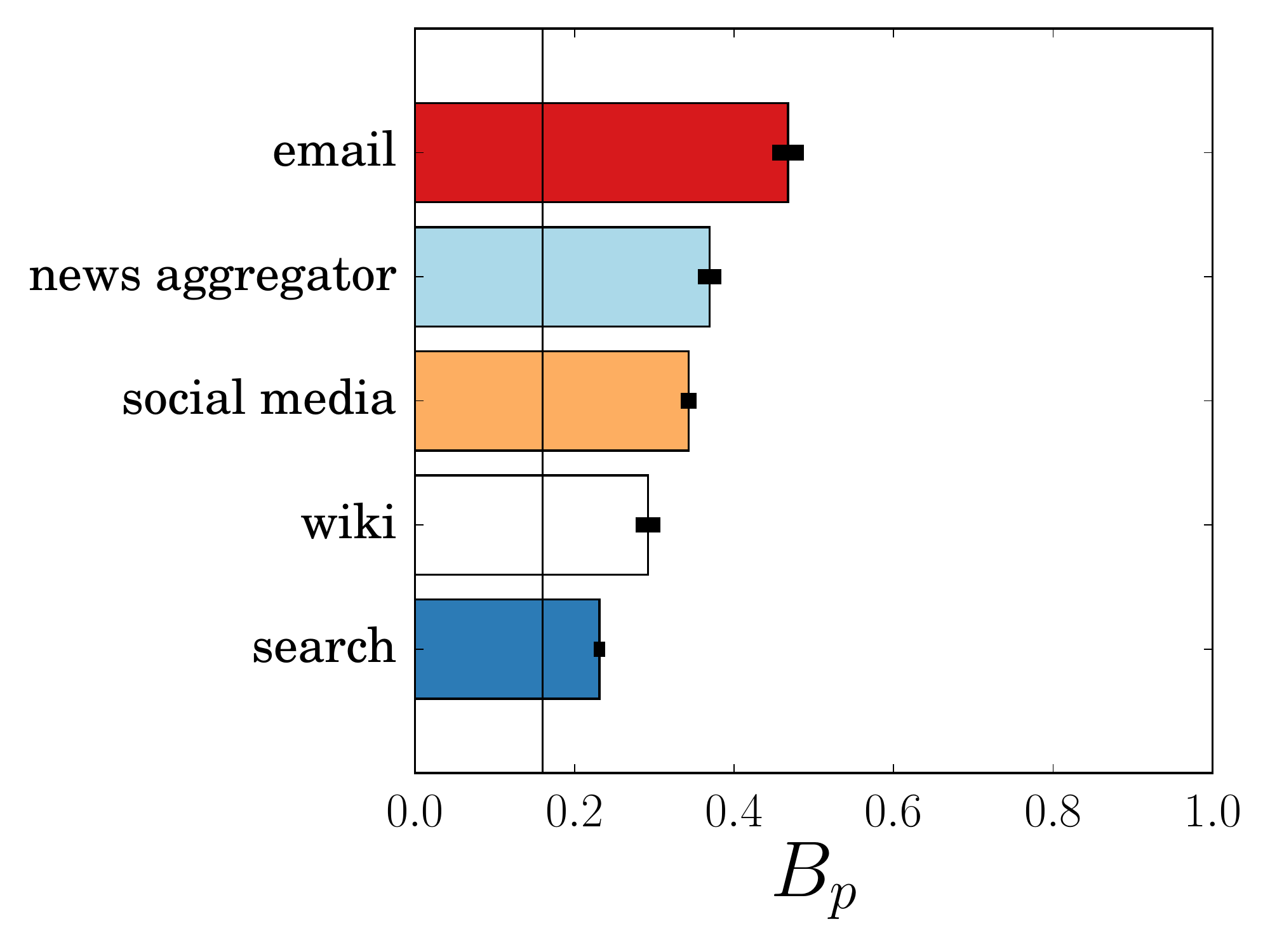} &
	\includegraphics[width=0.49\textwidth]{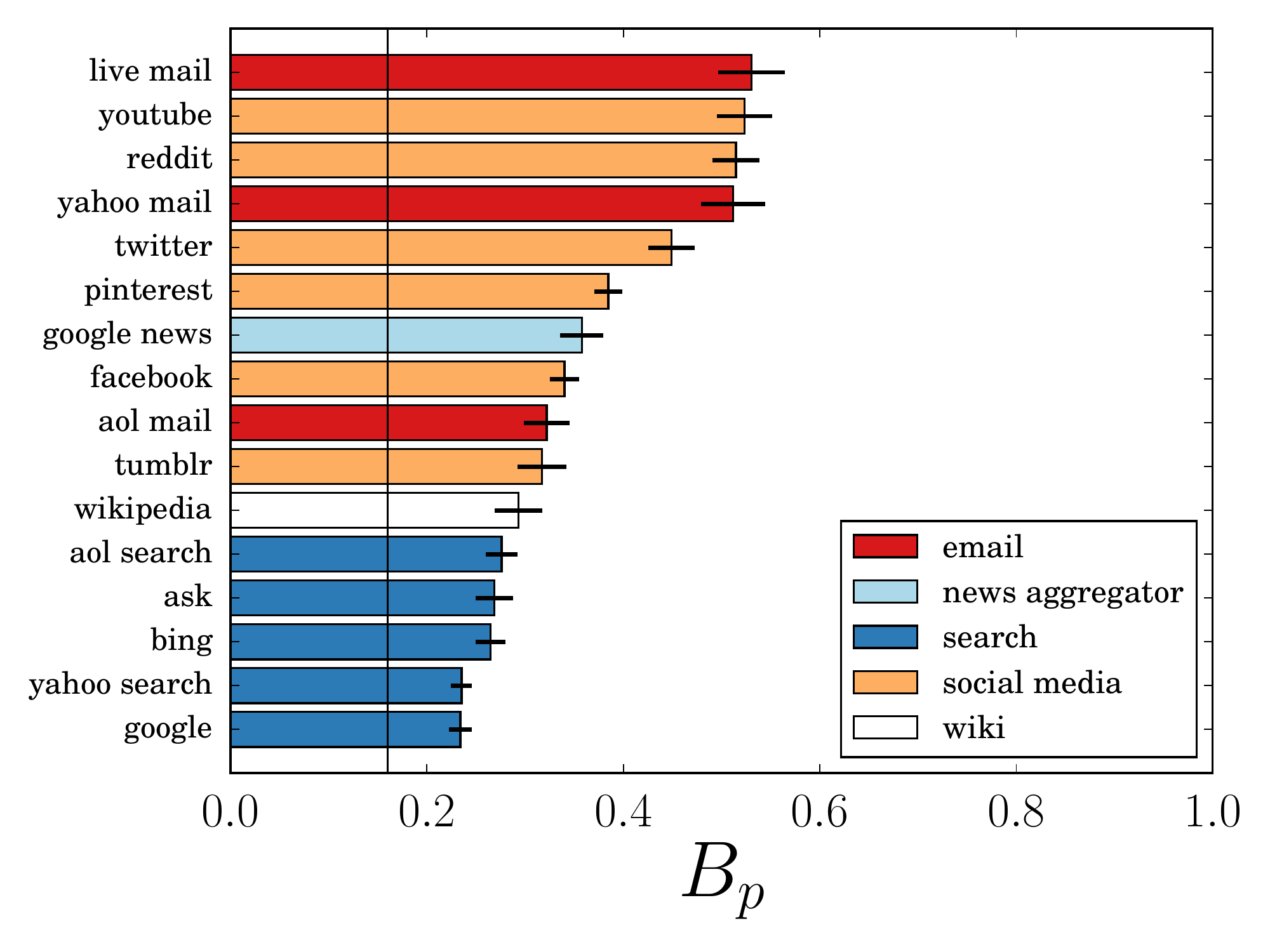}
	
\end{tabular}
\caption{Summary of popularity bias, averaged across users for (A) aggregate categories of interest and (B) individual domains. Error bars represent $\pm 2$ standard errors. The vertical lines show the baseline bias of a random walker through the Web domain network.}
\label{fig:pbias}
\end{figure}

One possible interpretation of the increase in popularity bias since 2006 is the higher proportion of traffic from social media. Thus, it is desirable to compare bias experienced by individual users in social media, search, and other popular online platforms. Figure~\ref{fig:pbias}A summarizes the popularity bias results obtained using the Gini coefficient-based definition of popularity bias introduced in ``Methods.'' We observe that email applications have the highest popularity bias, while search engines have the lowest. Google News, social media and Wikipedia are in the middle. To understand these results better, we drill down to the popularity bias of individual platforms, as shown in Figure~\ref{fig:pbias}B. We observe that search engines all have low popularity bias, providing evidence against H3 and extending the conclusions of \cite{Fortunato} from the aggregate to the user level.

We can also observe that email and social media display a greater variation, which is consistent with H1 and not surprising given the different functions and user interfaces of the social media systems under consideration. Facebook is one of the social media platforms with the least popularity bias. Its dominance in the social media traffic data may explain the moderate bias of social media in the aggregate. Also consistent with H1, all platforms display a higher popularity bias than expected from the baseline.

\begin{figure}[t]
\centering
\includegraphics[width=0.6\textwidth]{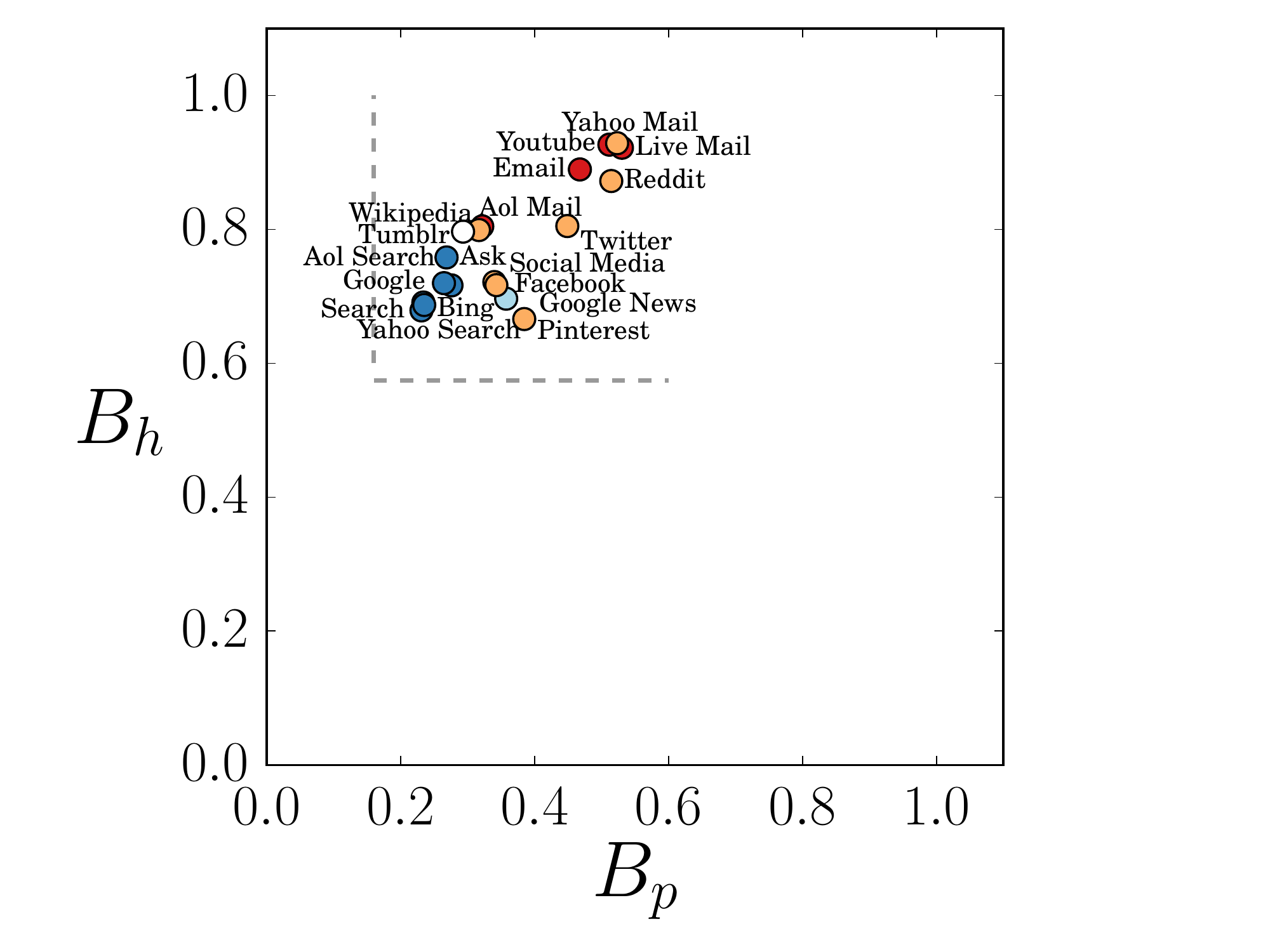} \\
\caption{Homogeneity versus popularity bias for applications of interest. The two types of bias are strongly correlated (Pearson's $r=0.83$, $p < 0.01$). We also show points corresponding to the aggregate traffic for email, social media and search. The dashed lines show the baseline biases of a random walker through the Web domain network.}
\label{fig:bias-map}
\end{figure}

Interestingly, we observe a strong correlation between the homogeneity and popularity bias of the domains of interest, as seen in Figure~\ref{fig:bias-map}. Because of the structure of the Web graph, we would expect to observe some correlation --- a random walker will visit domains with higher PageRank more frequently, which will lead to a skewed distribution of clicks, and thus, some homogeneity bias. It is an interesting question whether this correlation holds across more than the handful of domains analyzed in this paper. The skewed distribution of user traffic accross domains prevents us from answering this question here. As we will see below, the correlation is not observed when restricting the set of targets to news-only domains.

\subsection{User Analysis}
\label{sec:user-analysis}

In the previous analyses we sampled users separately for each application. This is necessary because the high variations in traffic volume across applications imply an insufficient overlap for a single sample of users engaged with all applications. To show that our results are robust to this methodology, we sampled a single set of users with sufficient traffic in each of the high-volume categories (search, social media, and email), when aggregated across applications. This sample includes 250,000 users having at least 1,000 clicks from each of the categories. We then measured popularity and homogeneity bias as a function of the mix of clicks for each user from the different categories.

\begin{figure}[t]
\centering
\begin{tabular}{ccc}
A (Density) & B (Popularity Bias) & C (Homogeneity Bias)\\
\includegraphics[width=0.33\textwidth]{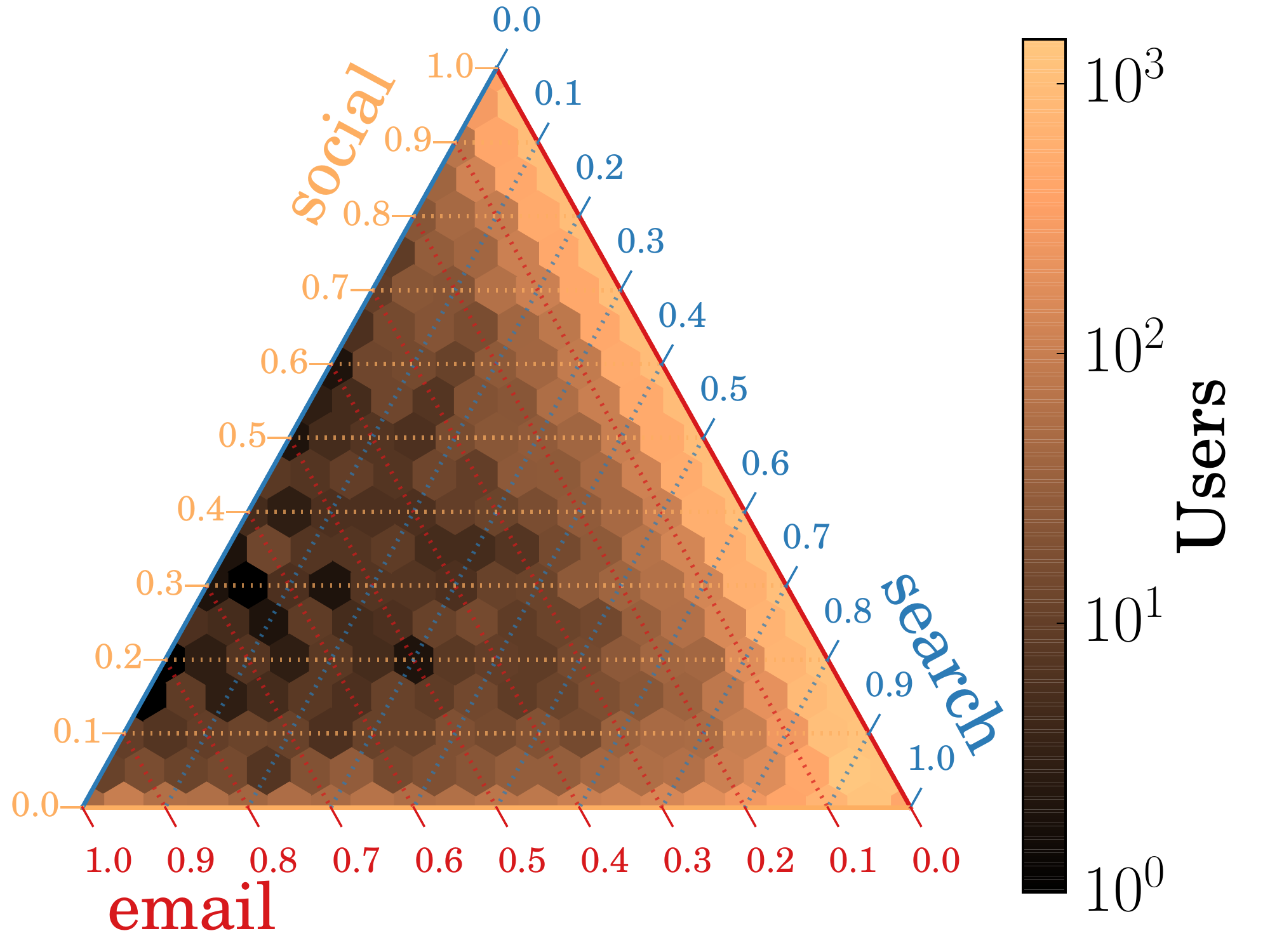} & 
\includegraphics[width=0.33\textwidth]{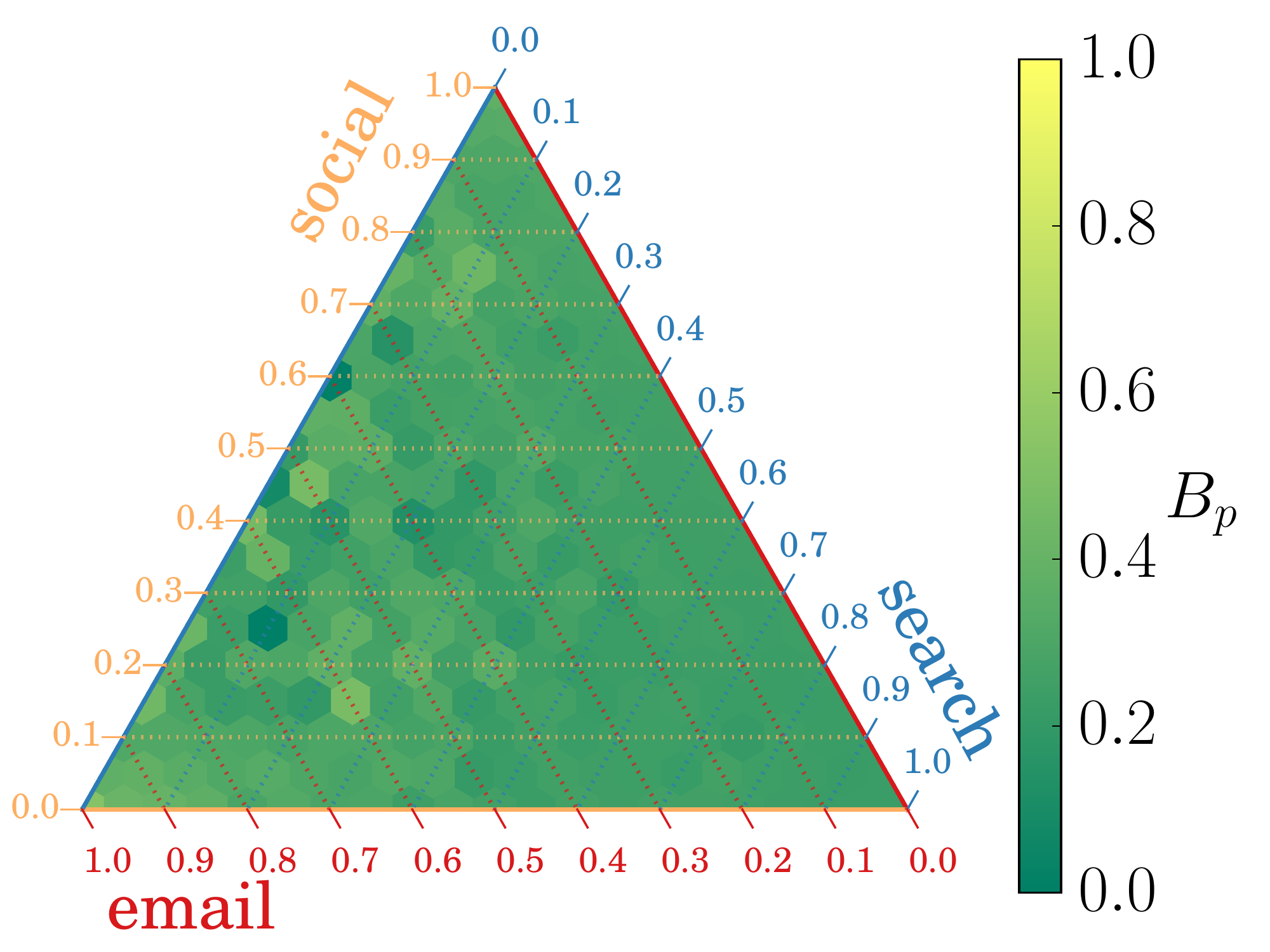} &
\includegraphics[width=0.33\textwidth]{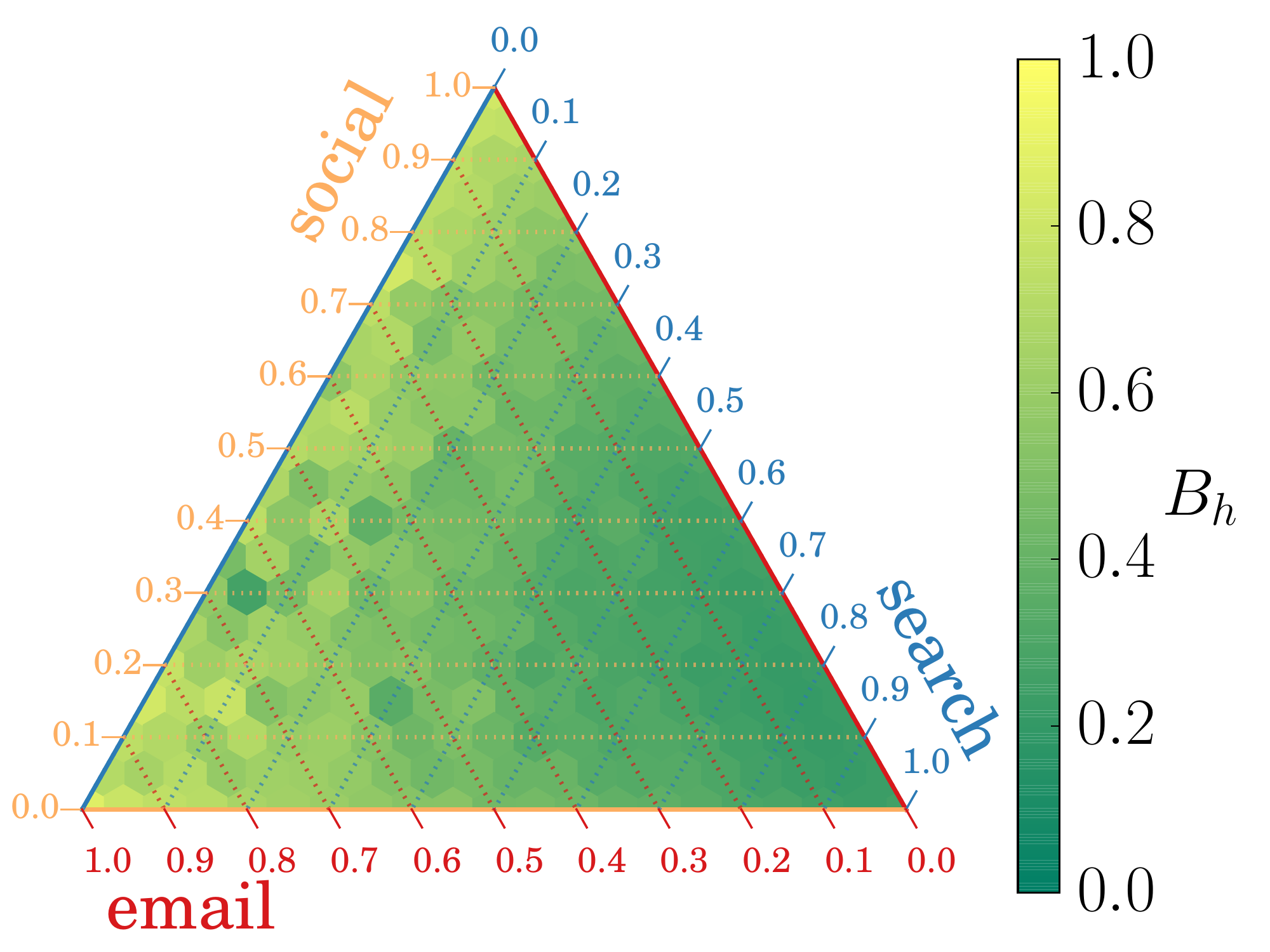}
\end{tabular}
\caption{Maps of users (A), their popularity bias (B), and their homogeneity bias (C). Users are mapped along three axes based on their mixes of social media, search, and email activity. Each hexagon is a bin containing the users with the corresponding proportion of clicks. The color of the hexagon corresponds to the density (A) or average bias (B, C) of the users in the bin.}
\label{fig:bias-user}
\end{figure}

Even among the three categories, the number of clicks differs by orders of magnitude. As we see in Figure~\ref{fig:bias-user}A, the vast majority of users are concentrated in the regions corresponding to low email and high search usage (the right-hand side and the bottom-right vertex of the triangle). We observe that users who search more (bottom-right vertex of Figures~\ref{fig:bias-user}B and~\ref{fig:bias-user}C) have the least homogeneity bias, followed by social media users, followed by email users. This is consistent with the earlier results, showing they are robust to our sampling method.

\subsection{News Analysis}

\begin{figure}[t]
\begin{tabular}{ccc}
A & B & C\\
	\includegraphics[width=0.33\textwidth]{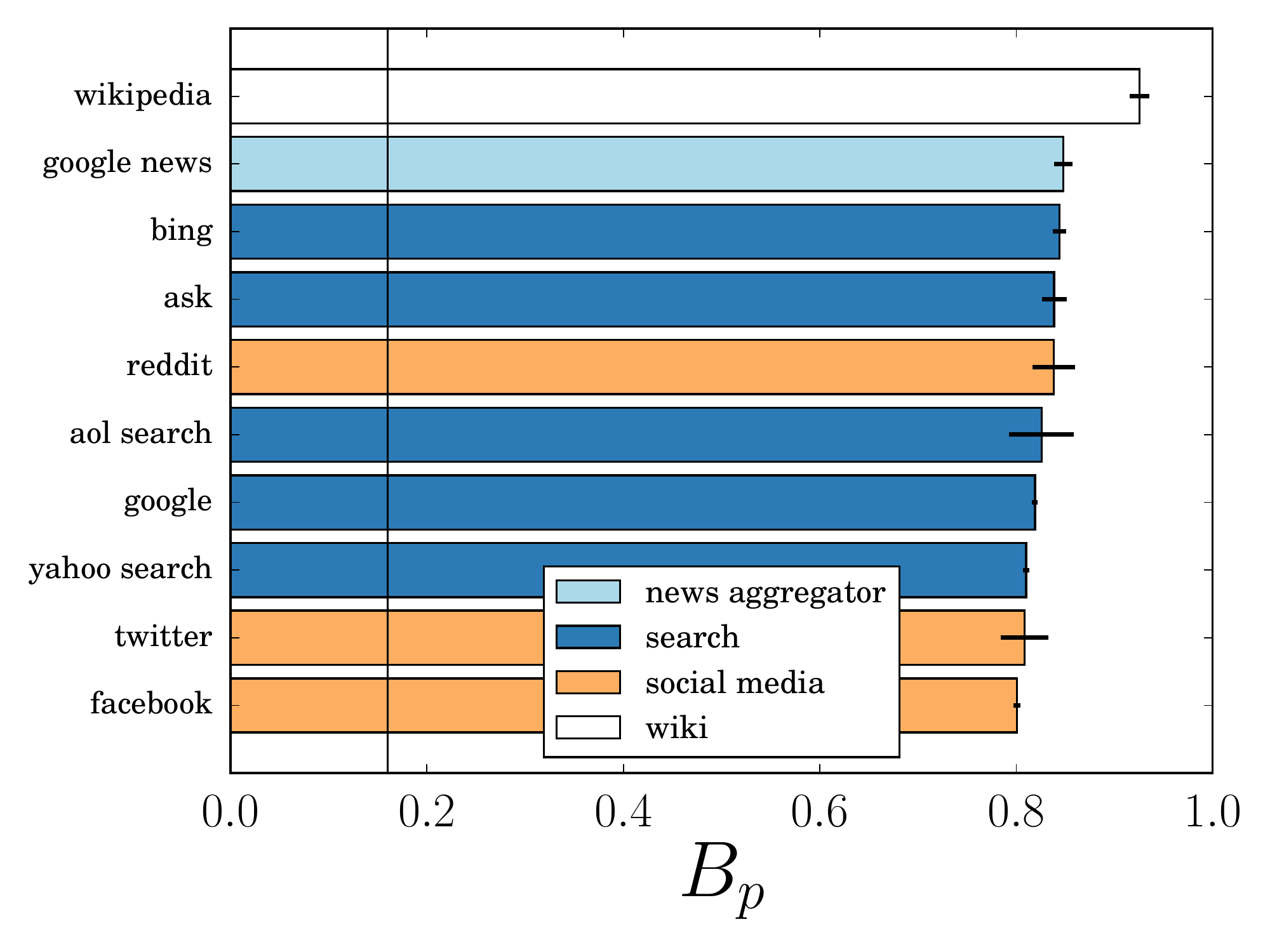} &
	\includegraphics[width=0.33\textwidth]{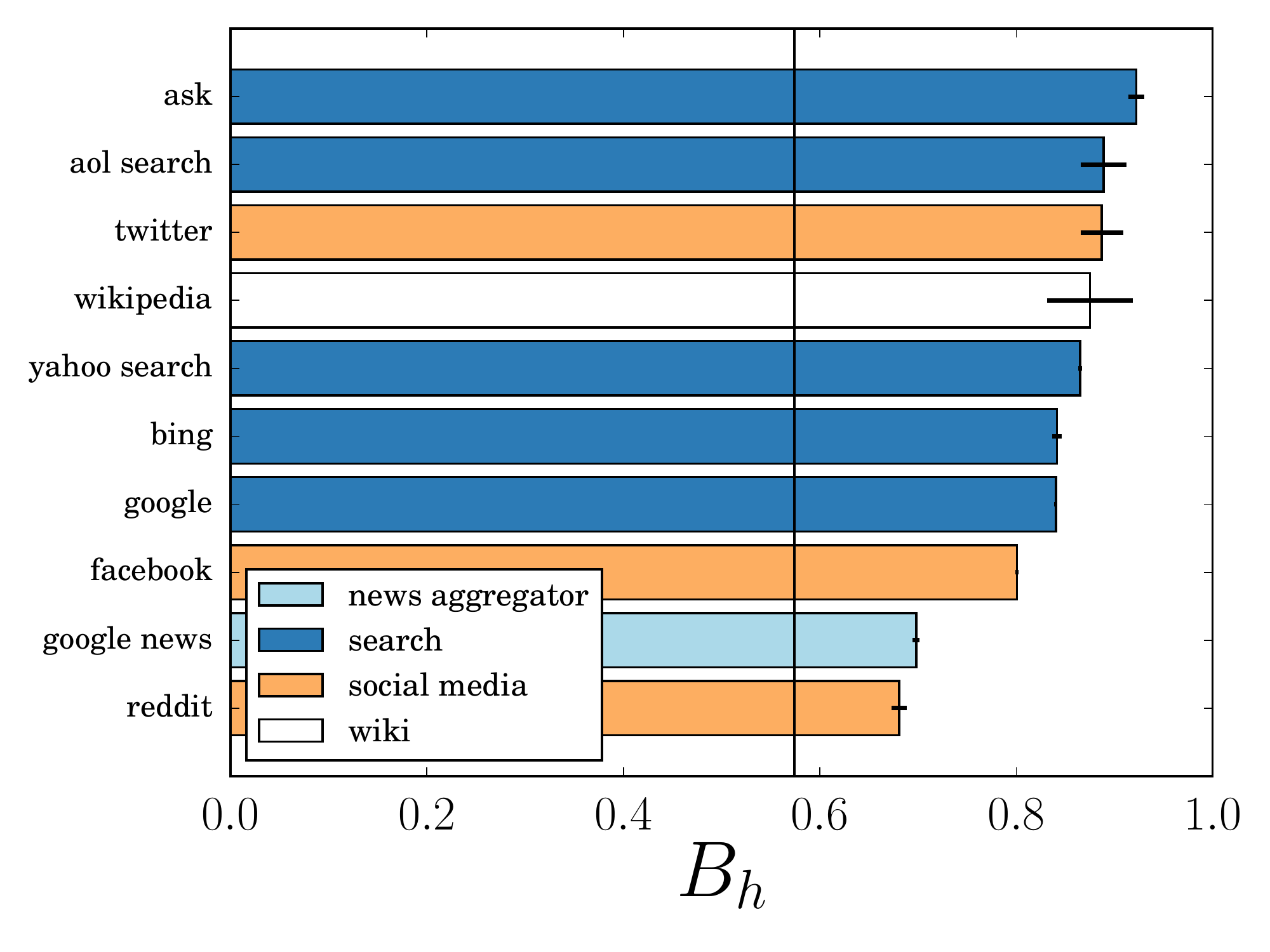} &
	\includegraphics[width=0.33\textwidth]{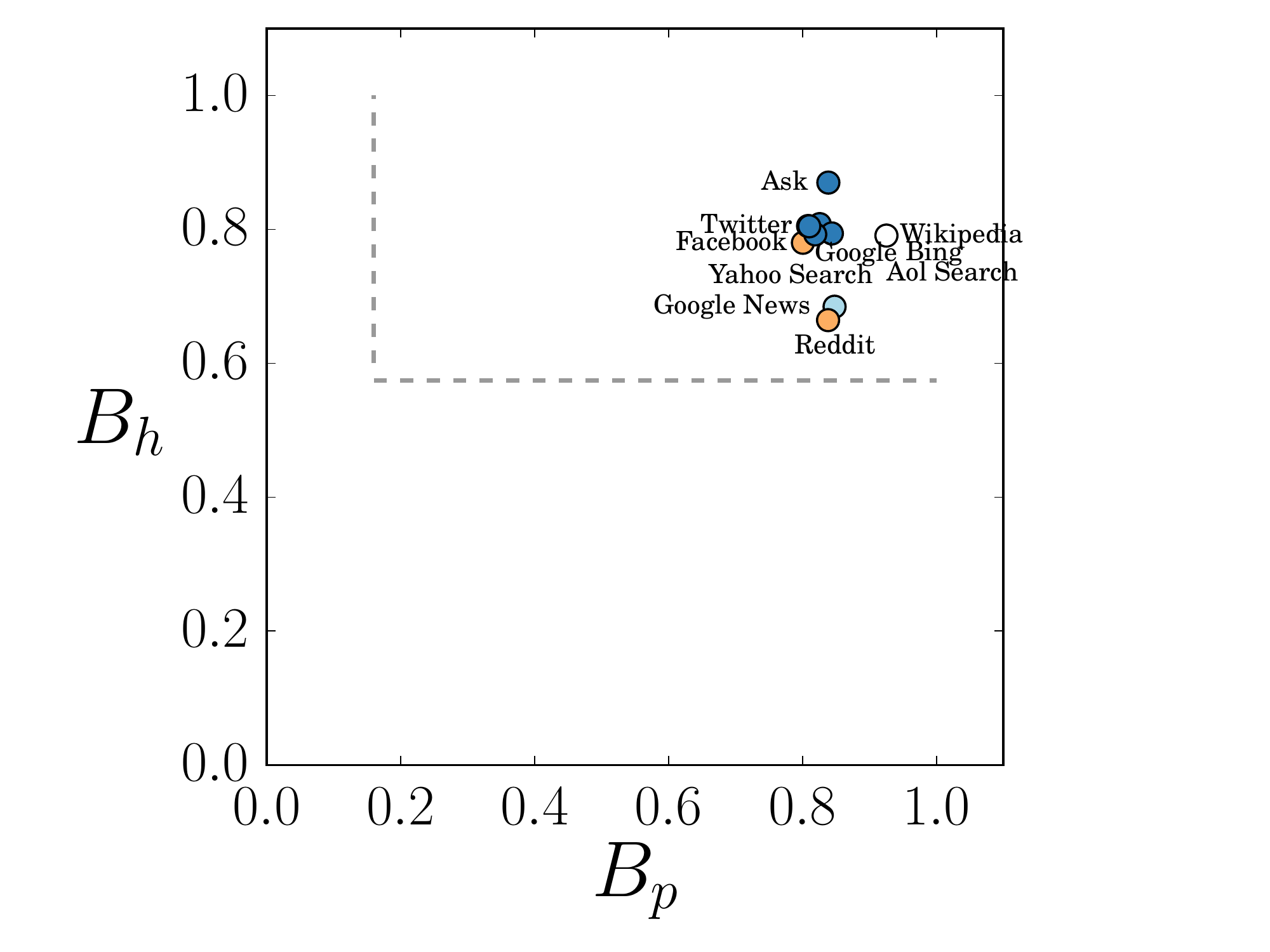}
\end{tabular}
\caption{(A) Popularity and (B) homogeneity bias for news targets. The vertical lines show the baseline biases of a random walker through the Web domain network. (C) The two types of bias are uncorrelated. The dashed lines show the baseline biases.}
\label{fig:news-targets}
\end{figure}

The role of platforms in the selective exposure of users to news is attracting increasing attention~\citep{BakshyAdamic,Lazer2018Science}. We therefore repeated the bias analysis for a subset of the data in which only news targets are considered. As shown in Figure~\ref{fig:news-targets}, we find a significant increase in the amount of popularity bias compared to the previous analysis considering all targets. This suggests that news traffic is dominated by popular outlets to a significantly higher extent than general traffic.
% to do: check how Hb was computer for news targets; which of the next sentences is true?
%This is not surprising, since the news dataset consists of news sites on a variety of topics like politics, sports, and entertainment, which are of interest to large populations and therefore have high PageRank compared to the full set of targets considered in the previous subsections.
We also observe that news aggregators like Google News and Reddit have the lowest homogeneity bias, suggesting that their users are exposed to a diverse set of news sources.
This is surprising given features on both sites that can contribute to a more biased exposure. For example, on Reddit, research
suggests that the number of votes on a story and thus its visibility can be effectively manipulated~\citep{glenski2016}; and on Google News, personalization algorithms based on user histories and extensive customization options for preferred topics and sources make it easy for users to filter undesirable content. Both of these systems can benefit from further analysis on what mechanisms help them achieve lower homogeneity bias.

\section{Discussion}
\label{sec:discussion}

In this paper we took a high-level view of bias on the Web. Instead of
examining specific mechanisms for information exposure that differ
from platform to platform, we focused on visits to Web domains. This
allowed us to quantify the bias of a large set of platforms in a
content-independent way. We defined two distinct
measures to capture popularity and homogeneity bias, which are robust
to different normalizations (see Appendix). Our results reveal a few
important insights and present a number of opportunities for further
analysis.

First, we saw significant agreement in the bias exhibited by distinct email and
search platforms, and significant differences among social media
platforms. This finding can be explained by the fact that, compared to
social platforms, email and search applications are similar to each other in their
functions and interfaces. Social media
platforms, on the other hand, can differ greatly in their focus: from
the shared content, such as videos or photos; to how much interaction
they foster among members; to whether they are geared toward
connecting with existing offline friends versus making new connections
with similar people online; and to the demographics of their users. This diversity of form and function in social media is reflected
in the differences in bias across platforms. It underscores a need to
investigate the mechanisms of specific platforms to which these
differences can be attributed.

Second, we found that search platforms are consistently less biased
than social media and email platforms. These findings are robust to
different normalizations and different sampling methods. At the same
time, we saw that when it comes to news exposure, the differences
between platforms tend to be relatively small. Analyzing news exposure
is important to the question of how online behavior affects societal
discourse. Our analysis could be extended with data about the
political bias of different news sources
\citep{BakshyAdamic}. Understanding how our measures of
exposure bias differ from the political bias of news sources can tell
us how good they are at capturing political
polarization. Investigation of a more narrow dataset, such as shares
of political news articles, could reveal other kinds of bias. An important caveat of our results, which can also
be addressed by such an analysis, is that we examined the traffic to
different domains but did not consider content in any
way. Naturally, some target domains are more diverse in the content
they provide than others. The homogeneity and popularity bias metrics
defined here could be extended to account for such differences by
weighing the targets according to content heterogeneity.

Third, both bias definitions can be used on any type of co-occurrence
data. We analyzed domain visits --- co-occurrences of users and
domains from clicks --- but our methods can be easily extended to links
shared on networks such as Facebook or Twitter, or topic discussions,
such as hashtags. Computing homogeneity and popularity bias
on sharing data from social media and other platforms will add another dimension
to the analysis --- the social network of users --- which does not
exist in the traffic data examined here.

Fourth, homogeneity and popularity bias are related, but they are not
measuring the same thing. Popularity bias implies homogeneity bias,
but the reverse is not true; an algorithm could in theory focus on
unpopular sites, thus achieving high homogeneity bias and low
popularity bias. The results in
Figures~\ref{fig:bias-map}~and~\ref{fig:news-targets}C suggest that a
correlation exists between popularity and homogeneity bias, but it is
dependent on restrictions we apply to the set of traffic sources and targets in
the data; the correlation disappears when focusing on news targets, or when considering many
sources rather than just the top platforms (not shown). The presence of a
correlation (or anti-correlation) between homogeneity and popularity
bias may be useful in identifying classes of sites, or even abuse. For
example, a spam site might have high homogeneity and low popularity
bias.

Fifth, we saw an increase in popularity bias compared to 2006 when using the methodology of \cite{Fortunato}. The higher use of social media could be one explanation for
this. As we saw, social media is consistently more biased than search,
the other category of traffic dominating the Web today. The increase in popularity bias
could also be attributed to other factors, such as browsing on mobile
devices, which offer smaller screens and an altogether different
browsing experience, and/or the changing demographics of users. More
work is needed to investigate these and other factors affecting
popularity bias.

Finally, exposure bias can lead to the formation of echo chambers that
foster the propagation of misinformation, such as hoaxes and fake
news. With increased focus on such activity~\citep{Lazer2018Science}
and the emergence of services for its tracking~\citep{hoaxy2016}, we
can examine whether users or communities characterized by high bias
are more likely to spread such misinformation.

\section{Conclusions}
\label{sec:conclusion}

In this paper, we presented findings on three important hypotheses. First, we have found support for hypothesis H1: popularity and homogeneity bias exist in all Web activities, but can significantly differ depending on the platform one is using to find or browse information. 

Second, while we found exposure biases in all platforms and categories, social media and search tend to be the primary channels through which users consume new information. We find support for hypothesis H2: social media tend to exhibit more homogeneity bias (as well as popularity bias) compared to search engines. This is consistent with our previous findings that social media may contribute to the emergence of ``social bubbles''~\citep{Nikolov}. However, individual social media platforms differ significantly from each other, suggesting that it is better to examine them separately. Indeed, while the aggregate traffic is dominated by Facebook, different platforms display varying degrees of bias; YouTube has significantly more homogeneity bias than Pinterest, and Twitter has significantly greater popularity bias than Tumblr; Facebook itself introduces relatively little bias compared to other social media platforms, but more bias compared to search engines.

Finally , we showed that across different Web activities, in the aggregate there exists a popularity bias consistent with the rich-get-richer structure of the Web. However, this bias is not significantly stronger in search engines, as previously theorized. Our analysis does not provide support for hypothesis H3: search engines do not bias exposure towards popular websites more than other platforms.

The differences across distinct applications almost disappear when we focus on traffic toward news sites. In this case all platforms are highly biased in terms of both homogeneity and popularity, with small differences between them. Google News and Reddit are the exceptions, with lower homogeneity bias. 

The two types of bias are not completely independent; higher $B_p$ necessarily implies higher $B_h$. It is somewhat surprising that no correlation is observed for traffic to news sites, suggesting the need to examine this relationship further by modeling, or by imposing fewer restrictions on the source domains. 

To further validate the present findings, it is important for our analyses to be reproduced for other user populations and time periods; the relationships we have found may change over time as a result of both user behaviors and platform algorithms. 

The analyses presented here allow for the first time to compare exposure biases across different techno-social systems. The bias metrics defined in this paper are broadly applicable to any type of co-occurrence data beyond Web traffic. Examples include sharing, liking, or commenting on hashtags, news articles, and products. Our methods can be supplemented with content and/or network data to formulate additional bias metrics, thus helping inform the future design of Web platforms as they face challenges such as disinformation and manipulation. 
 
\section{Acknowledgements}
\label{sec:ack}

We are grateful to Ricardo Baeza-Yates for useful discussions about traffic normalization. In addition, we thank Mark Meiss for his work on analysis of Web traffic from Indiana University, which served as an inspiration for some of the work in this paper. Finally, thanks to Yahoo Research for making data available for research and for hosting F.M. during his sabbatical; part of this work was performed during this visit.

\appendix
\section{Traffic Normalization}
\label{sec:appendix}

When computing bias, it is important to normalize the clicks in such a way that they represent comparable exposure for each user. In the main text, this normalization is done by sampling the same number of clicks per user, thus equating comparable exposure between users with the same amount of effort. This approach has the advantage that it discounts volume effects on the bias measurements. However, not all users are equally active, so we could be comparing an hour's worth of activity by one user to ten minutes' worth of activity by another.

To see how sensitive the bias measures are to these different formulations of comparable exposure, we collected all clicks by users active in two separate time periods --- the 5 days from Feb 1 to Feb 5 2015, and the 31 days in October 2014. All users who made at least ten clicks were included in these datasets.

\begin{figure}
\centering
\begin{tabular}{cc}
    \multicolumn{2}{c}{February 1-5, 2015 (Popularity Bias)} \\
    \includegraphics[width=0.49\textwidth]{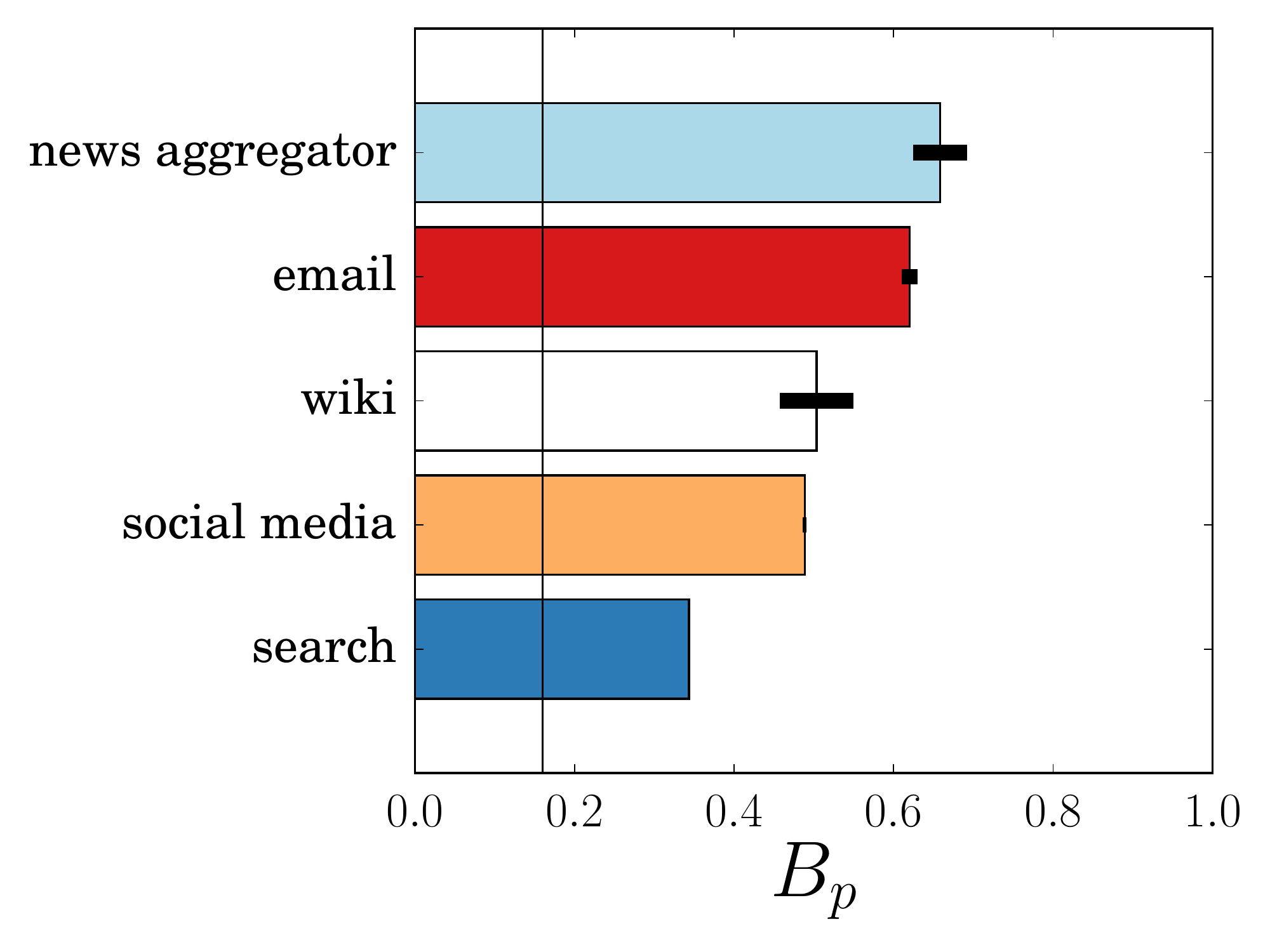} &
    \includegraphics[width=0.49\textwidth]{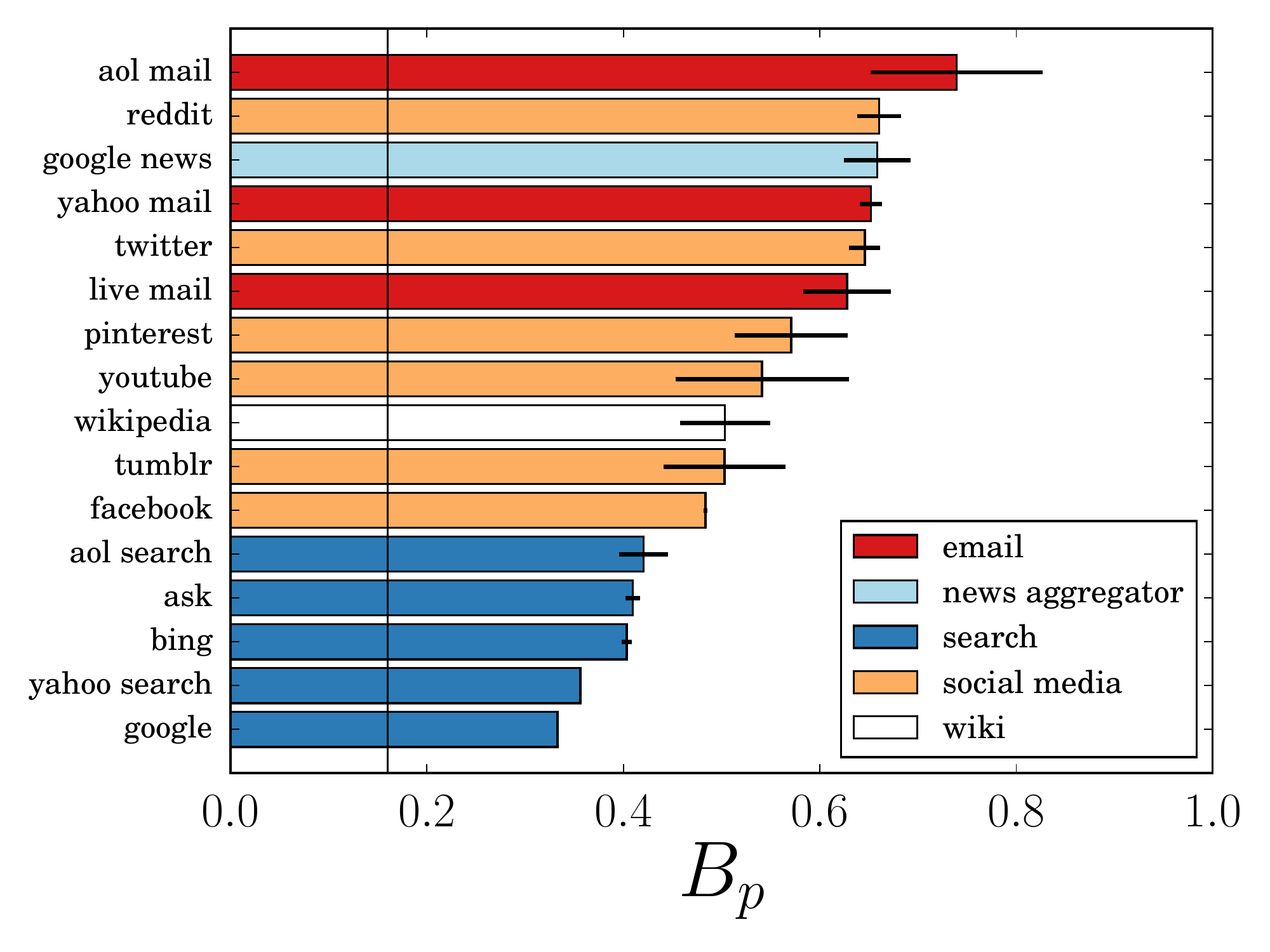} \\
    \multicolumn{2}{c}{February 1-5, 2015 (Homogeneity Bias)} \\
    \includegraphics[width=0.49\textwidth]{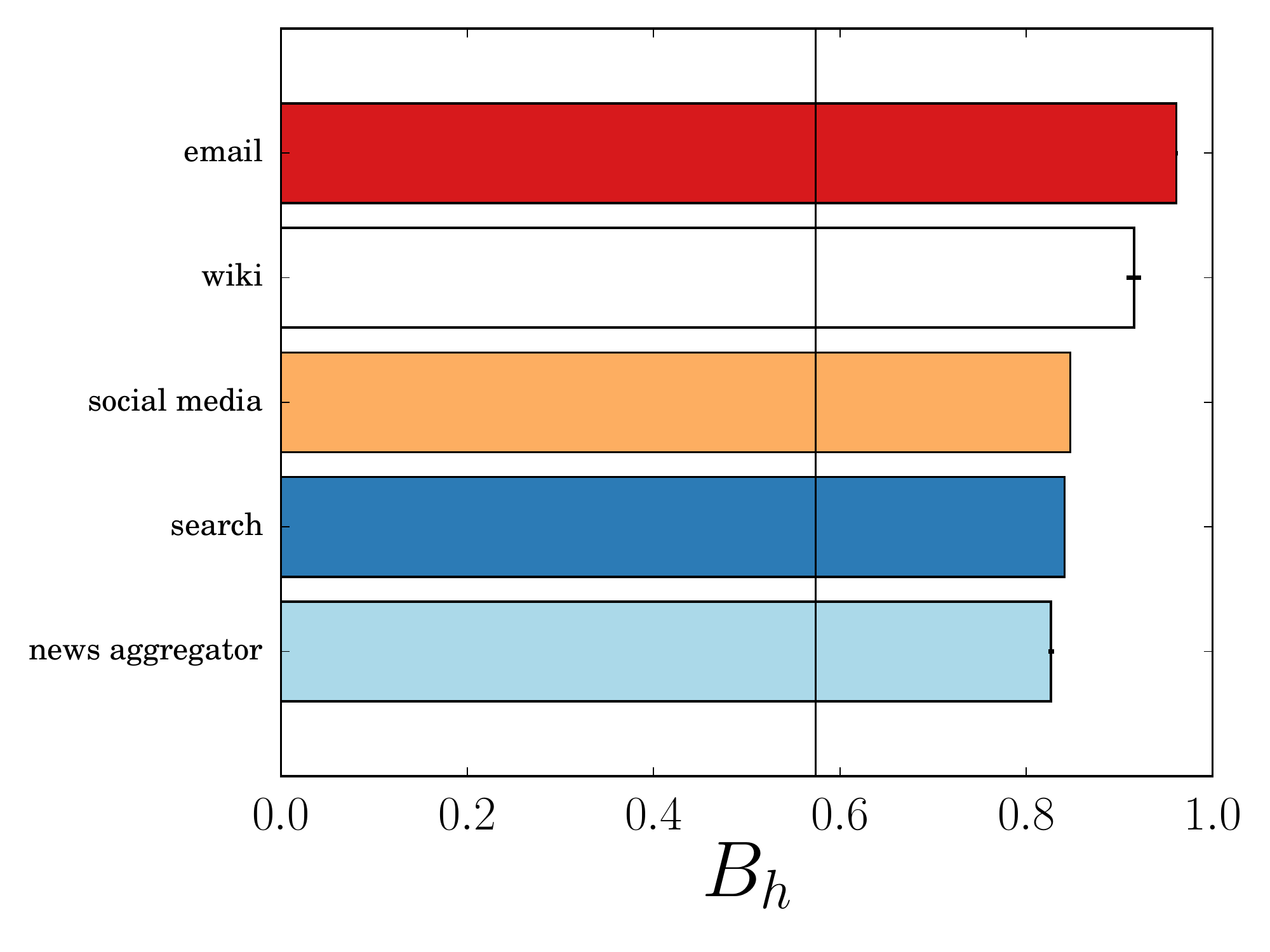} &
    \includegraphics[width=0.49\textwidth]{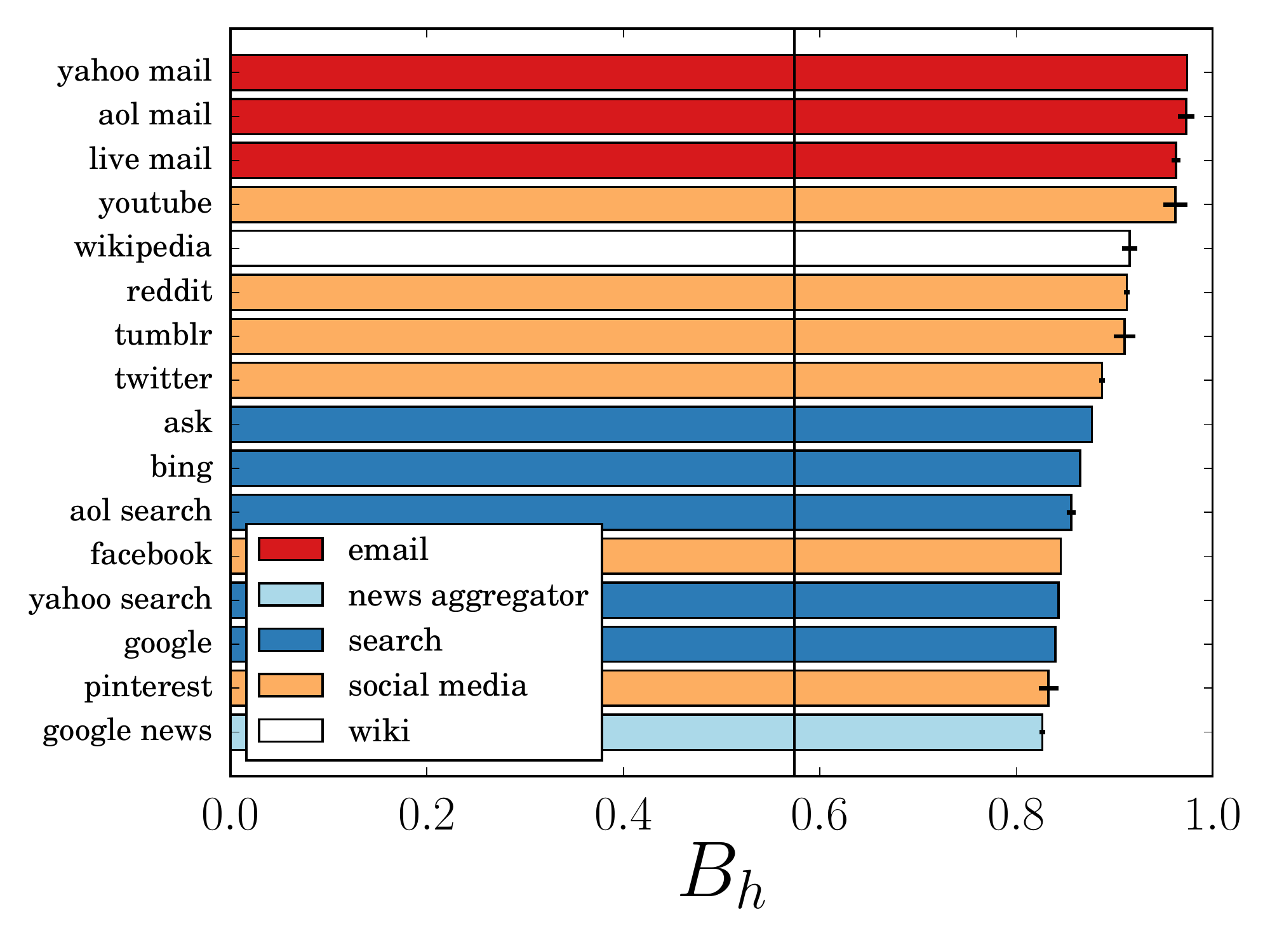} \\    
  \end{tabular}
  \caption{Popularity (top) and homogeneity (bottom) bias for all clicks between Feb 1, 2015 and Feb 5, 2015 by users with at least 10 clicks. The error bars are $+/-$ two standard errors. The vertical lines show the baseline biases of a random walker through the Web domain network.}
  \label{fig:feb2015-01to05}
\end{figure}

\begin{figure}
\centering
\begin{tabular}{cc}
    \multicolumn{2}{c}{October 1-31, 2014 (Popularity Bias)} \\
    \includegraphics[width=0.49\textwidth]{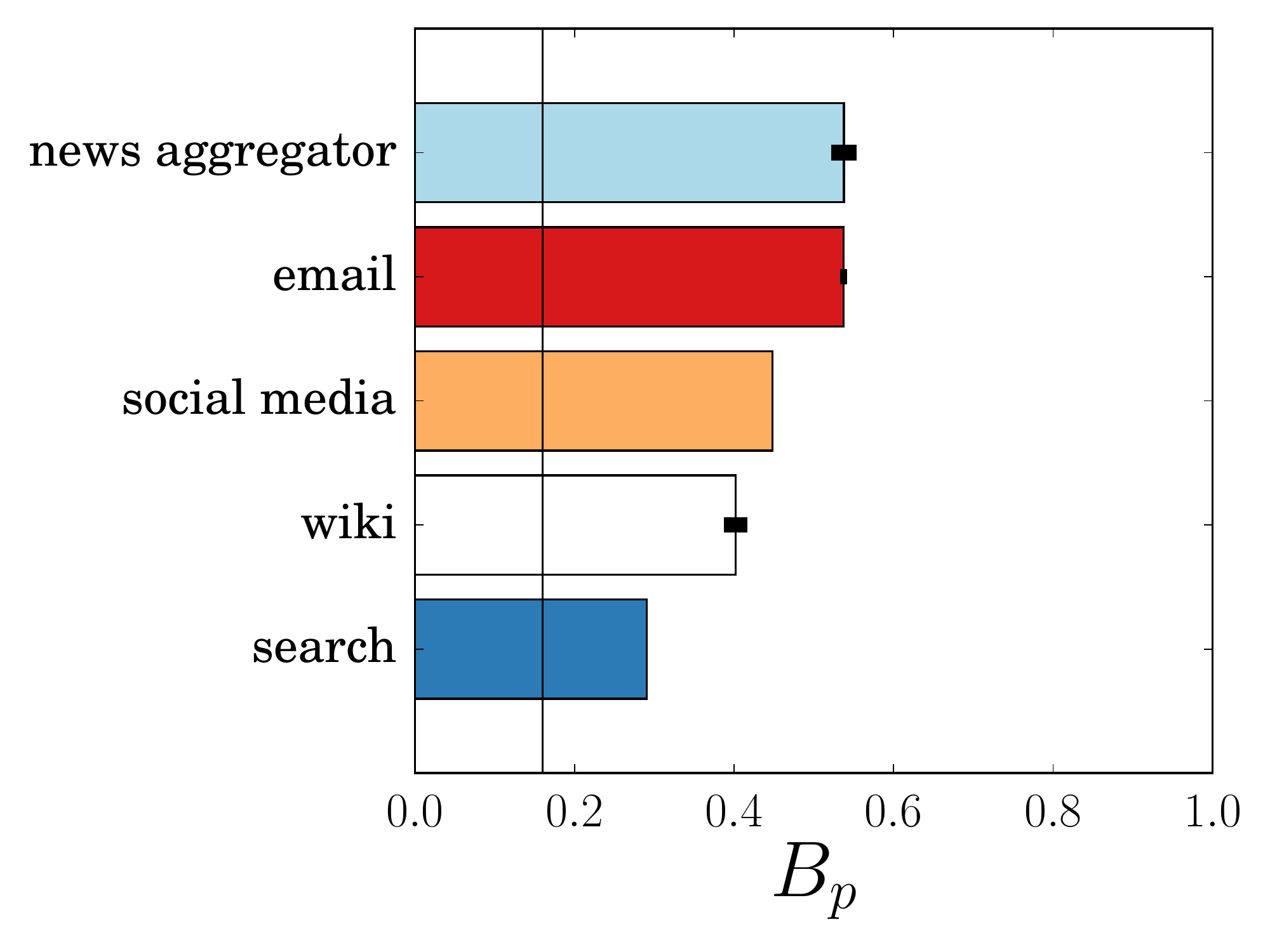} &
    \includegraphics[width=0.49\textwidth]{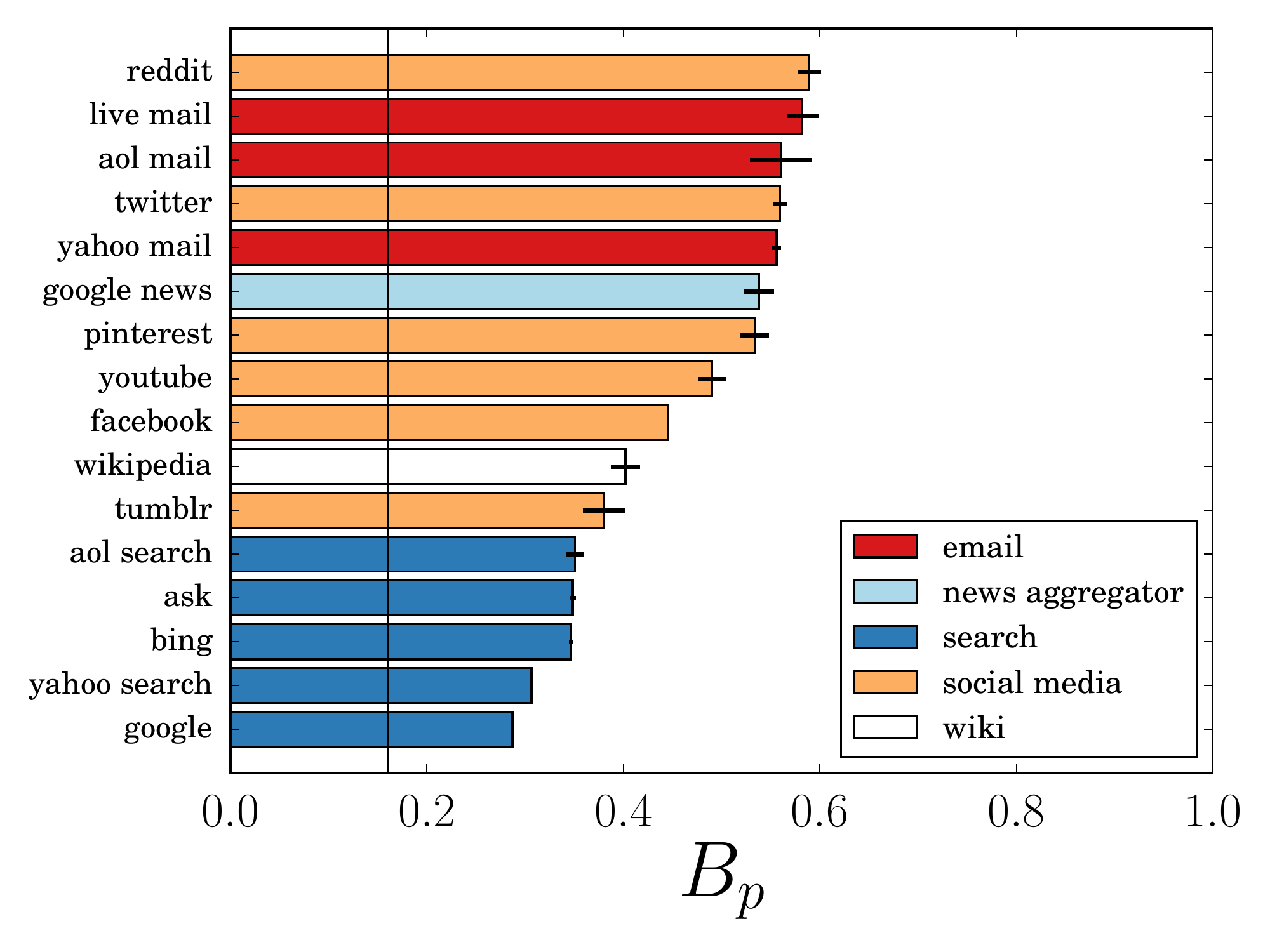} \\
    \multicolumn{2}{c}{October 1-31, 2014 (Homogeneity Bias)} \\
    \includegraphics[width=0.49\textwidth]{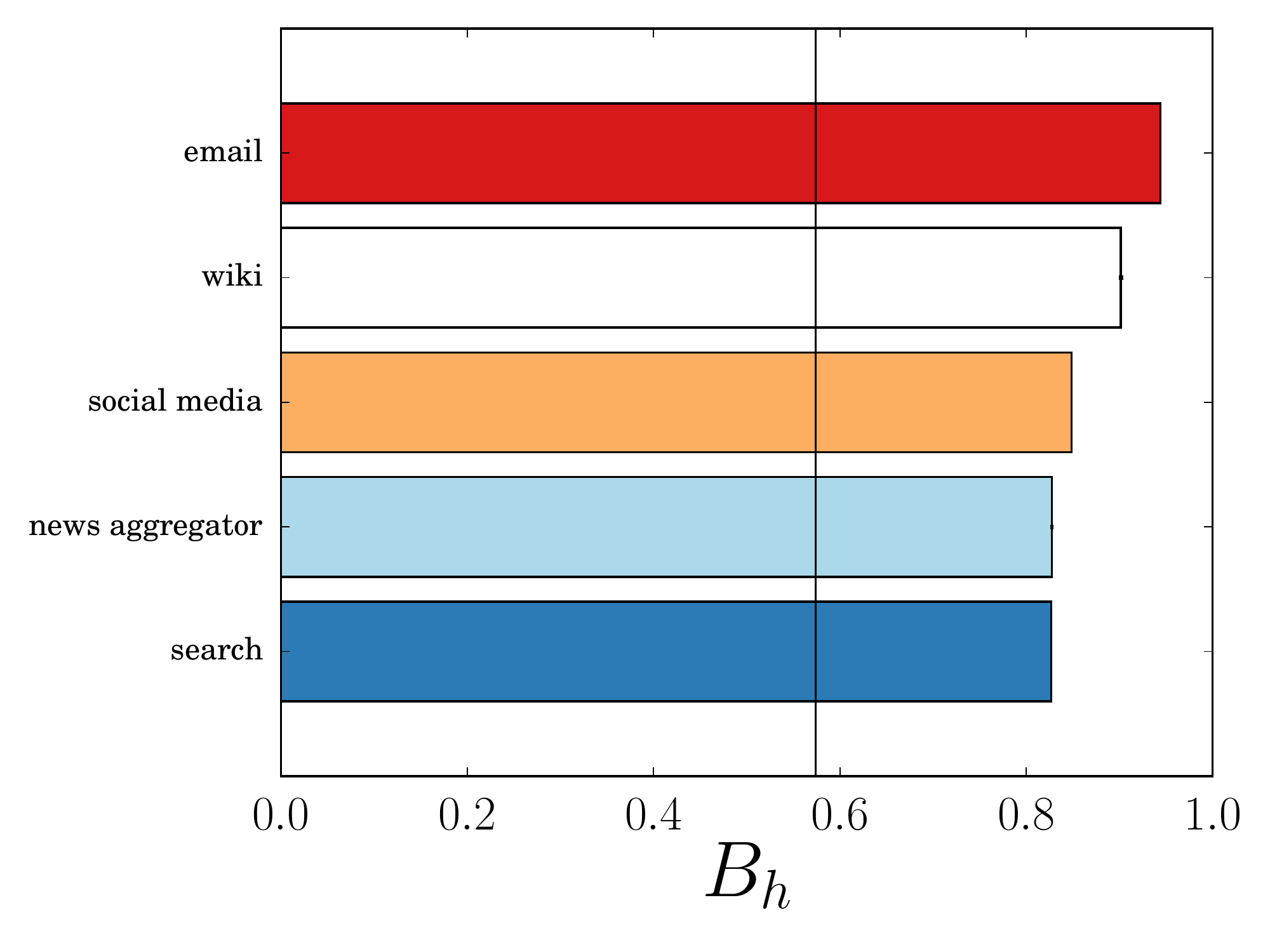} &
    \includegraphics[width=0.49\textwidth]{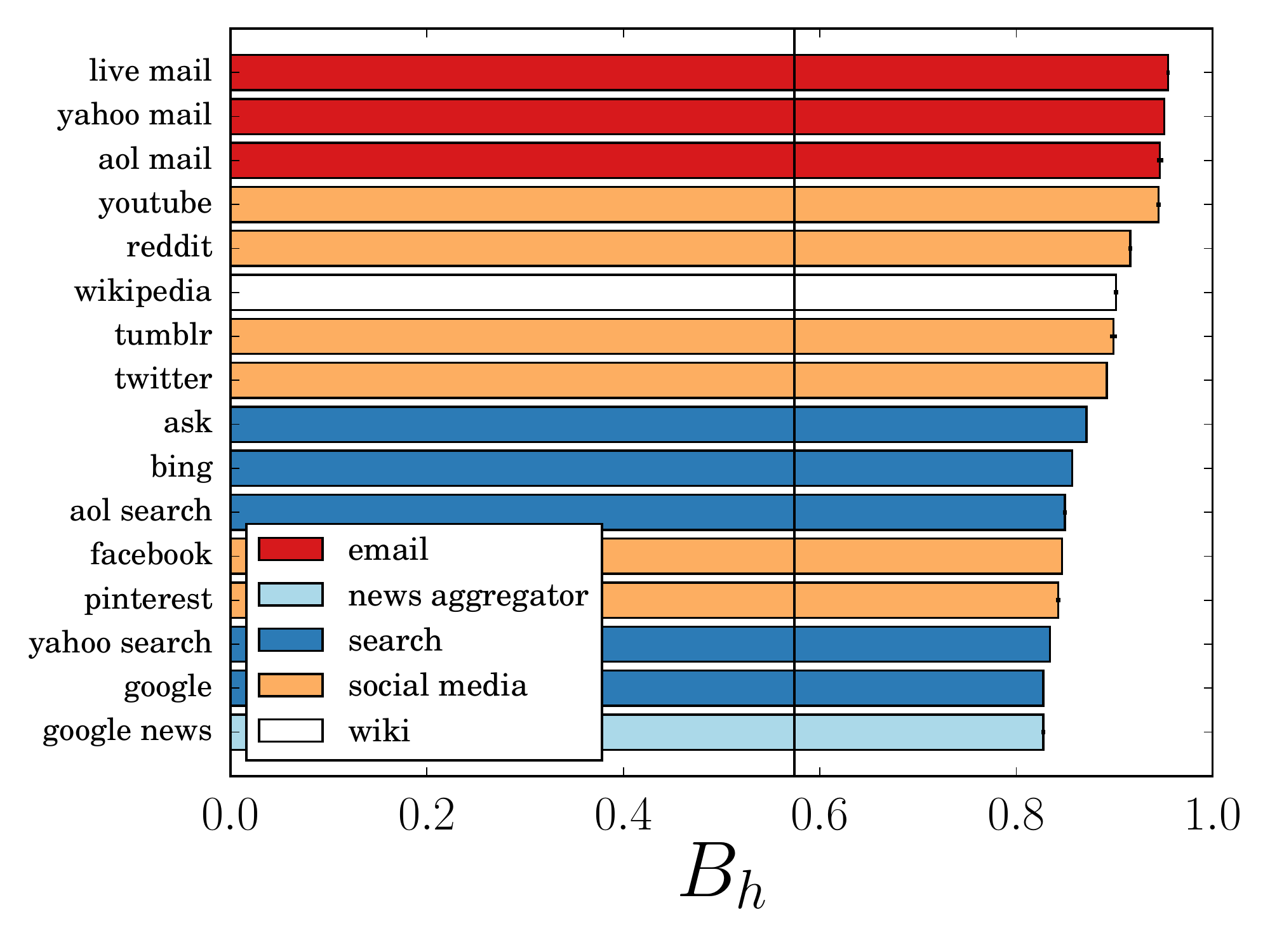} \\    
  \end{tabular}
  \caption{Popularity (top) and homogeneity (bottom) bias for all clicks between Oct 1, 2014 and Oct 31, 2014 by users with at least 10 clicks. The error bars are $+/-$ two standard errors. The vertical lines show the baseline biases of a random walker through the Web domain network.}
  \label{fig:oct2014}
\end{figure}

We measured the bias for both time periods and present the results in Figures~\ref{fig:feb2015-01to05} and \ref{fig:oct2014}. We observe consistency between the bias measured during these different time periods. In addition, these results are mostly consistent with those shown in the main text of the paper. We conclude that the bias measurements are fairly robust with respect to how we normalize traffic for user exposure.

\bibliography{refs}

\end{document}